\documentclass[aps,pre, preprint, groupedaddress,superscriptaddress,showpacs,preprintnumbers,amsmath,amssymb]{revtex4}

\usepackage{amsmath}
\usepackage{amssymb}
\usepackage{amsthm}
\usepackage{verbatim}

\usepackage{graphicx}
\usepackage{dcolumn}
\usepackage{bm}
\begin{document}

 \newcommand{\dth}{\ensuremath{\delta\theta}}
\newcommand{\dk}{\ensuremath{\delta\kappa}}
\newcommand{\ka}{\ensuremath{\kappa}}
\newcommand{\mrben}{\ensuremath{\mathrm{bend}}}
\newcommand{\mrpr}{\ensuremath{\mathrm{pres}}}
\newcommand{\dd}{\ensuremath{\mathrm{d}}}
\newcommand{\la}{\ensuremath{\langle}}
\newcommand{\ra}{\ensuremath{\rangle}}

\title{Collapse and folding of pressurized rings in two dimensions}
\author{Eleni Katifori}\email{ekatifori@mail.rockefeller.edu}
\affiliation{Department of Physics, Harvard University, Cambridge, Massachussetts, 02138}
\affiliation{Center for Studies in Physics and Biology, Rockefeller University, New York, New York, 10065}
\author{Silas Alben}
\affiliation{SEAS, Harvard University, Cambridge, Massachussetts, 02138}
\affiliation{School of Mathematics, Georgia Institute of Technology, Atlanta, Georgia, 30332}
\author{David R. Nelson}\email{nelson@physics.harvard.edu}
\affiliation{Department of Physics, Harvard University, Cambridge, Massachussetts, 02138}

\date{\today}

\begin{abstract}
Hydrostatically pressurized circular rings confined to two dimensions (or cylinders constrained to have only $z$-independent deformations) undergo Euler type buckling when the outside pressure exceeds a critical value. We perform a stability analysis of rings with arc-length dependent bending moduli and determine how weakened bending modulus segments affect the buckling critical pressure. Rings with a 4-fold symmetric modulation are particularly susceptible to collapse. In addition we study the initial post-buckling stages of the pressurized rings to determine possible ring folding patterns.

\end{abstract}
\maketitle

\section{\label{sec:Introduction}Introduction }

Stiff elastic rings confined to two dimensions and subjected to in plane forces, such as a hydrostatic pressure, have been extensively studied in the past, mainly in the context of the Euler-Bernoulli elastica theory and bifurcations. Related problems arise for cylinders, aligned with the $z$-axis, and constrained to have only $z$-independent deformations. The buckling instability of such a system under uniform pressure has been approached as an eigenvalue problem and issues such as existence and stability of solutions have been discussed in several works \cite{Levy84, Carrier47, Tadjbakhsh67, Chaskalovic95, Atanackovic98,Wang1985,Wang1993}. In addition to the problem's appeal from a purely academic perspective, there are physical systems where such studies find direct applicability. One example is carbon nanotubes, whose mechanical properties under pressure have been investigated with the ultimate goal of manufacturing microscopic pressure gauges \cite{Zang04}.

The study of stiff tubes subject to forces finds biological applications in animal veins that collapse under certain loads and in the marginal band of non mammalian erythrocytes and thrombocytes and mammalian platelets. The marginal band is a ring-shaped microtubular bundle at the periphery of the erythrocyte or thrombocyte that provides support to the cell membrane and helps it withstand capillary flow stresses~\cite{Joseph84, Waugh89, Italiano03}. The structural and mechanical properties of the marginal band have not been fully understood yet, but it has been claimed that non-uniformity in the spontaneous curvature might play a critical role in determining the flattened ellipsoidal cell shape \cite{Cohen98}.

Rings under forces warrant an extensive study not only because of the above applications, but also because they provide some intuition about their three dimensional counterparts. In fact, planar pressurized thermally fluctuating rings with finite bending rigidity have been used as simplified models of fluctuating vesicles such as red blood cells\cite{Leibler89}. In particular, Leibler, Singh and Fisher (LSF) have considered thermal fluctuating closed planar chains, including both pressure and curvature energy terms\cite{Leibler87}. Different aspects of the LSF closed 2-d random walk problem in the inflated and deflated regime have been discussed in the literature, using a number of analytical and computational techniques (see e.g. \cite{Levinson92, Romero92, Marconi93, Gaspari93A, Haleva06, Mitra07}).

In most fluctuating closed chain work, the elastic properties of the chain are assumed constant and uniform. Heterogeneities in the LSF model have been considered in the form of impurities  and intercalated molecules as an additional degree of freedom in the equations (see e.g. \cite{Morikawa95} and references therein). However, planar pressurized rings with quenched inhomogeneities that manifest themselves in arc-length dependent elastic properties appear to have received little attention in the literature.    

In this work, using a combination of analytic and computational techniques, we study the stability of inhomogeneous, inextensible, planar pressurized rings with finite spontaneous curvature at zero temperature. For large rings, the approximation of inextensibility is always appropriate (see below). As we shall see, spontaneous curvature becomes especially interesting when rings with regions of weakened elastic strength are considered, even in the limit of zero pressure. In Sec.~\ref{sec:Buckling} we briefly review the stability analysis of uniform pressurized rings \cite{ Carrier47, Tadjbakhsh67}, and introduce formalism that will be used later in the paper. In Sec.~\ref{sec:Spontcurv} we consider the effects of spontaneous curvature on unpressurized rings with nonuniformities. In Sec.~\ref{sec:Criticalpress} we derive the critical pressure for different ring bending modulus profiles and discuss the aspects of the bending modulus variability that contribute most to the stability of the ring. The most pronounced weakening arises in rings with a 4-fold symmetric modulation of the bending rigidity. Last, in Sec.~\ref{sec:Collapse}, we explore the initial stages of the post-buckling, non-uniform ring behavior \footnote{The thermal behavior of inhomogeneously elastic rings will be discussed in a future paper (E. Katifori and D. R. Nelson, to be published).}. In the Appendices we include derivations omitted from the main part of the article and discuss the ring buckling problem in the context of second order phase transitions.

\section{\label{sec:Buckling}Buckling of uniform rings with a spontaneous curvature}

For completeness, in this Section we briefly review the buckling of uniform pressurized rings, as discussed in Refs.~\cite{Carrier47, Tadjbakhsh67}, and calculate the critical pressure \footnote{For a treatment of the buckling transition of \textit{extensible} rings the interested reader can refer to \cite{Atanackovic98}.}. The bending modulus $\kappa$ of a rod scales as the fourth power of the thickness $h$, $\kappa\sim h^4$; whereas the stretching modulus as the second power \cite{Landau70}. From these facts it can be shown that, for rings that are sufficiently large, i.e. $h\ll R$, the energetic cost of bending is always smaller than the stretching energy
\footnote{Consider, for example, a rod of thickness $h$ and length $L$ with compressive forces applied at the ends, sufficient to produce a strain of order unity. If $Y$ is the three dimensional Young's modulus of the rod, the energy of compression is of order $Yh^2L$. However, if the rod buckles out of the axis of compression (thus trading compression energy for bending energy), the energy is now of order $Y h^4/L$. Bending deformations will be preferred whenever $Eh^4/L \ll Eh^2L$, i.e., $(h/L)^2\ll 1$. More generally, we can consider rings or rods inextensional whenever their perimeter or length greatly exceed their thickness.}. 
Note that it is always possible to deform rods inextensibly, which is not the case for two-dimensional shells. In this work the rings that we consider are assumed inextensional.

\begin{figure}
\includegraphics[scale=0.25]{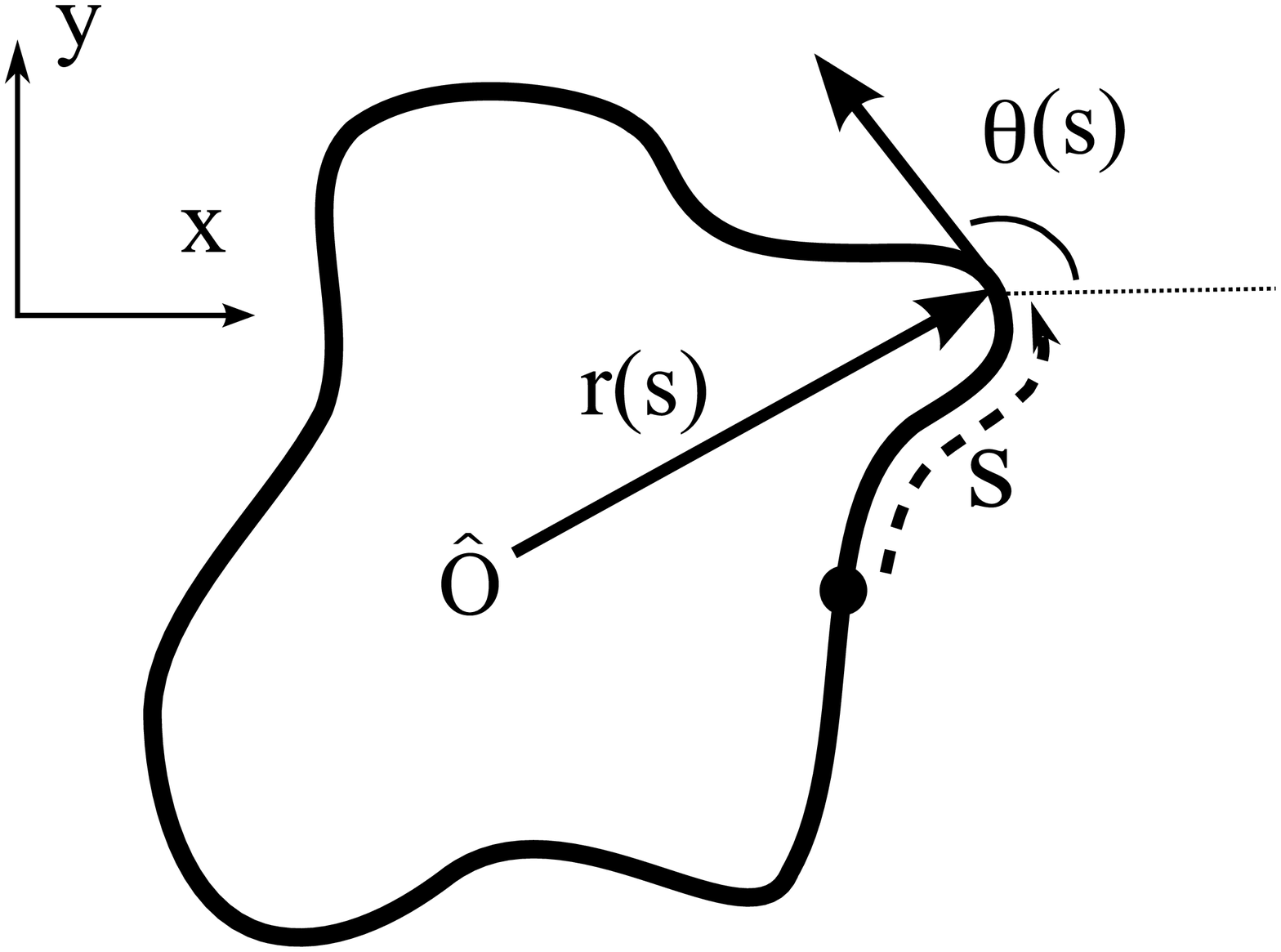}
\caption{\label{fig:stronggr} Parameterization of an inextensible ring. $\theta(s)$ is the angle between the tangent at arc-length $s$ and the $x$-axis.  }
\end{figure}

Upon ignoring torsion and spontaneous curvature, the bending energy of an inextensional circular ring of radius $r_c$, circular cross section and an arc-length dependent bending modulus $\kappa(s)$ reads \cite{Landau70}:
\begin{equation}\label{bendingenergym1}
E_{\mrben} = \frac{1}{2}\int\limits_{ - \pi r_c }^{\pi r_c}  {\dd s\; \kappa(s) \left[\mathbf{t}\times \frac{d\mathbf{t}}{ds}\right]}^2
= \frac{1}{2}\int\limits_{ - \pi r_c }^{\pi r_c}  {\dd s\; \kappa(s) \left[\frac{d^2\mathbf{r}}{ds^2}\right]}^2
\end{equation}
where $\mathbf{t}\equiv\frac{\dd\mathbf{r}}{\dd s} $ is the tangent of the ring at a point $\mathbf{r}(s)=[x(s), y(s)]$ parametrized by the arc-length $s$, as shown in Fig.~\ref{fig:stronggr}. The ring is assumed inextensible, so in the arc-length parametrization, $| \frac{d\mathbf{r}}{ds} |=1$.
The bending energy of a ring of spontaneous curvature $c_0$ (which in general is $c_0\ne 1/r_c$)   reads:
\begin{equation}\label{bendingenergy1}
E_{\mrben} = \frac{1}{2}\int\limits_{ - \pi r_c }^{\pi r_c}  {\dd s\; \kappa(s) \left[c(s) - c_0\right]}^2
\end{equation}
where $c(s)= \hat{m} \cdot \left[ \mathbf{t}(s) \times  \frac{d\mathbf{t}(s)}{ds} \right] $ 
is the curvature of the ring at a point $\mathbf{r}(s)=[x(s), y(s)]$ parametrized by the arc-length $s$, as shown in Fig.~\ref{fig:stronggr}. . Here $\hat{m}$ is a unit vector perpendicular to the plane of the ring. The curvature $c(s)$ is given by the derivative of the unit tangent vector $\mathbf{t}(s)$ with respect to $s$. 
The work done on the system by exerting a pressure difference $p$ between the inside and the outside of the ring is:
\begin{eqnarray}\label{volumeenergy1}
E_{\mrpr} &=& p (A-\pi r_c^2)\nonumber\\
& =& \frac{p}{2}\int\limits_{ - \pi r_c }^{\pi r_c} \left[x(s) \mathrm{d}y(s)-y(s) \mathrm{d}x(s)\right]  -p\;\pi r_c^2
\end{eqnarray}
where $A$ is the area enclosed by the ring.

Despite the simple form assumed by the area when a Cartesian parameterization is used, this choice actually is not the most convenient one. In order to confine the functions described by $\mathbf{r}(s)=[x(s), y(s)]$ to a set that describes inextensional rings of perimeter $2\pi r_c$, we need to impose not only periodic boundary conditions $\mathbf{r}(- \pi r_c)=\mathbf{r}(\pi r_c)$ but also an inextensibility constraint  $| \frac{\dd\mathbf{r}}{\dd s} |=1$ for all $s$. This constraint is trivially satisfied if we recast $\mathbf{r}(s)=[x(s), y(s)]$ in terms of $\theta(s)$, the angle between the tangent vector and the $x$-axis (see Fig.~\ref{fig:stronggr}). The coordinates $x(s)$ and $y(s)$ then read:
\begin{eqnarray}\label{xandyofs}
x(s) & = & \int\limits_{ - \pi r_c}^s {\cos \left[ \theta (\xi)\right]  d\xi}\nonumber\\
y(s) & = &  \int\limits_{ - \pi r_c}^s {\sin \left[ \theta (\xi)\right]  d\xi}.
\end{eqnarray}
This choice simplifies Eq.~\eqref{bendingenergy1}, since the curvature can now be rewritten in terms of the tangent angle as $c(s)=\frac{\mathrm{d}\theta(s)}{\mathrm{d} s}$:
\begin{equation}\label{bendingenergy1b}
E_{\mrben} = \frac{1}{2}\int\limits_{ - \pi r_c }^{\pi r_c}  {\dd s \; \kappa(s) \left[\frac{\mathrm{d}\theta(s)}{\mathrm{d} s}- c_0\right]}^2.
\end{equation}
The pressure now couples to an area with an interesting nonlocal dependence on the field $\theta(s)$:
\begin{equation}\label{area}
A[\theta (s)] = \frac{1}{2}\int\limits_{-\pi r_c}^{\pi r_c} {\dd s} \int\limits_{-\pi r_c}^s {d\xi \sin \left[ {\theta (s) - \theta (\xi )} \right]}.
\end{equation}
Note than when $\kappa(s) = \kappa = \mathrm{const}$, the bending energy simplifies:
\begin{equation}\label{bendingenergy1bb}
E_{\mrben} = \frac{\ka}{2}\int\limits_{ - \pi r_c }^{\pi r_c}  {\dd s \left(\frac{\mathrm{d}\theta(s)}{\mathrm{d} s}\right)}^2 - 2\pi\kappa c_0 + \pi\kappa r_c c_0^2 ,
\end{equation}
so the spontaneous curvature only contributes an additive constant in this limit.
 For a given $p$ and $\kappa(s)$, the shape of the ring is determined by functional minimization of the total energy:
\begin{equation}\label{energyfunctional}
E[\theta(s)]=E_{\mrben}+E_{\mrpr}
\end{equation}
subject to the constraint
\begin{equation}\label{constraint1}
\int\limits_{-\pi r_c}^{\pi r_c} \dd s e^{i\theta(s)}=0,
\end{equation}
which ensures periodicity in $x(s)$ and $y(s)$, and
\begin{equation}\label{constraint2}
\theta(-\pi r_c)=\theta(\pi r_c)+2\pi
\end{equation}
which guarantees a complete $2\pi$ circuit of $\theta(s)$ as one moves around the ring.
A last condition, $\theta(-\pi r_c)=-\pi/2$, can be used to choose arbitrarily a solution from the group of degenerate solutions that are connected by rigid body rotations.

The equation for the curvature  $c(s)$, derived from minimizing the functional of Eq.~\eqref{energyfunctional} plus two Lagrange multiplier terms from the real and imaginary parts of Eq.~\eqref{constraint1}, is nonlinear and analytically challenging (see App.~\ref{appeqderiv}). For constant $\kappa$ (a uniform ring) it reads:
\begin{equation}\label{bucklingequation}
\kappa\left( \frac{\dd^2 c}{\dd s^2}+\frac{c^3}{2 } \right) -p -b c = 0
\end{equation}
where $b$ is an integration constant. Note that the spontaneous curvature $c_0$ does not appear explicitly when $\kappa$ is constant.
The problem however can be reformulated using an ansatz and reduced to solving a set of transcendental and algebraic equations \cite{Carrier47}. 
It is found that below a critical pressure equal to $p_c=3\kappa/r_c^3$, the ring retains its circular shape. However, as the pressure increases above $p_c$, the ring suffers an elliptical deformation, and ultimately assumes a peanut-like shape.

The critical pressure can also be obtained by linearizing Eq.~\eqref{bucklingequation}. The quantity $\Delta c(s) = c(s) -1/r_c$ is very small for an infinitesimally deformed ring, and for constant $\kappa$ the zero pressure shape is always circular. Hence we can rewrite Eq.~\eqref{bucklingequation} in terms of $\Delta c(s)$ and and keep only the lowest order terms:
\begin{equation}\label{bucklingequationlinear}
\frac{\dd^2 \Delta c}{\dd s^2}+\frac{1}{r_c^2}\left(\frac{3}{2}-\frac{b r_c^2}{\kappa} \right) \Delta c = -\frac{1}{2r_c^3}+\frac{p}{\kappa}+\frac{b}{r_c\kappa}.
\end{equation}
A particular solution of Eq.~\eqref{bucklingequationlinear} is a constant proportional to the right hand side of the equation. However since $\Delta c$ is periodic and integrates to zero, the right hand side of the equation must vanish. This constraint fixes the integration constant $b$:  $b=\frac{\kappa}{2r_c^2}-p r_c$. 

Eq.~\eqref{bucklingequationlinear} now reduces to the equation for a classical harmonic oscillator, where the arc-length plays the role of time.
The nontrivial solution  $\Delta c\sim \sin(\sqrt{1+\frac{p r_c^3}{\kappa}}\frac{s}{r_c}+\mathrm{const})$ only satisfies the periodic boundary conditions provided $\sqrt{1+\frac{pr_c^3}{\kappa}}=n$, where $n$ is an integer. The solution $n=1\rightarrow p=0$ does not satisfy Eq.~\eqref{constraint1}, and the next allowed solution ($n=2$) gives a critical pressure $p_c=3\kappa/r_c^3$.

The continuous rotational symmetry of the undeformed ring is broken at and beyond the buckling pressure, where the ring becomes ellipsoidal. This transition can be viewed as a continuous, second order phase transition, with pressure as the control parameter. A short discussion about the transition and the order parameter is included in Appendix~\ref{ap_phasetrans}.

The stability study in this section holds only for constant bending modulus and constant spontaneous curvature. The value of the bending modulus determines the rigidity of the ring and the critical buckling pressure $p_c$. However, provided that the bending modulus is constant, the spontaneous curvature does not affect the behavior of the pressurized ring, since the cross term $c_0$ in the quadratic bending energy density in Eq.~\eqref{bendingenergy1b} integrates to a $\theta(s)$-independent constant. Inextensible rings with $\textit{any}$ constant spontaneous curvature should have the same critical pressure and identical buckling behavior. As we shall see, the situation is quite different for nonuniform rings.

\section{\label{sec:Spontcurv}Spontaneous curvature effects in rings with nonuniformities with zero pressure}

As discussed at the end of Sec.~\ref{sec:Buckling}, spontaneous curvature plays no role in the buckling of uniform rings.
However, as is well known from everyday experience, this is no longer true for rings with varying bending rigidity with, for example, weakened or strengthened areas. Even in the absence of a pressure, if the spontaneous curvature of the ring is different than its natural circular curvature $1/r_c$, then the bending stresses inherent in the ring will be partially released as the ring assumes a deformed non-circular shape. In this section we discuss the $p = 0$ equilibrium shape of simple model rings with varying $\kappa$ and $c_0\ne 1/r_c$.

Henceforth, unless otherwise stated, to simplify the formalism we will measure all lengths in units of the circular radius $r_c$, which is equivalent to considering rings of total length $2\pi$.

\begin{figure}
\includegraphics[scale=0.45]{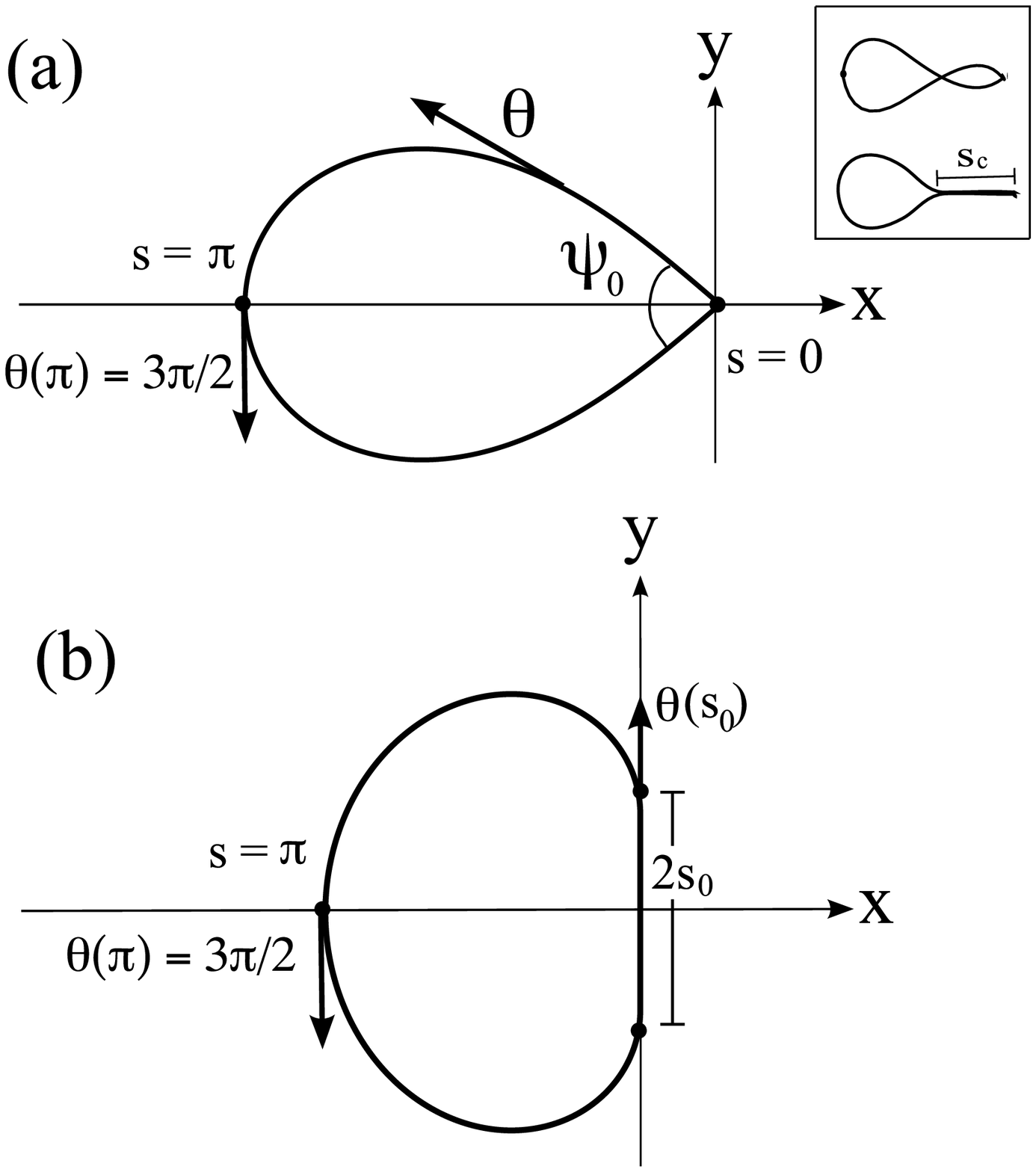}
\caption{\label{fig:explX} Parametrization of rings with non-uniformities. Inset: Hinged ring with $c_0\lesssim -1$. Top: Intersections allowed. Bottom: No intersections allowed. (a)Hinged ring. (b)Ring with strong sector.}
\end{figure}

\subsection{\label{sec:weak_e}Hinged rings}

\begin{figure}
\includegraphics[scale=0.6]{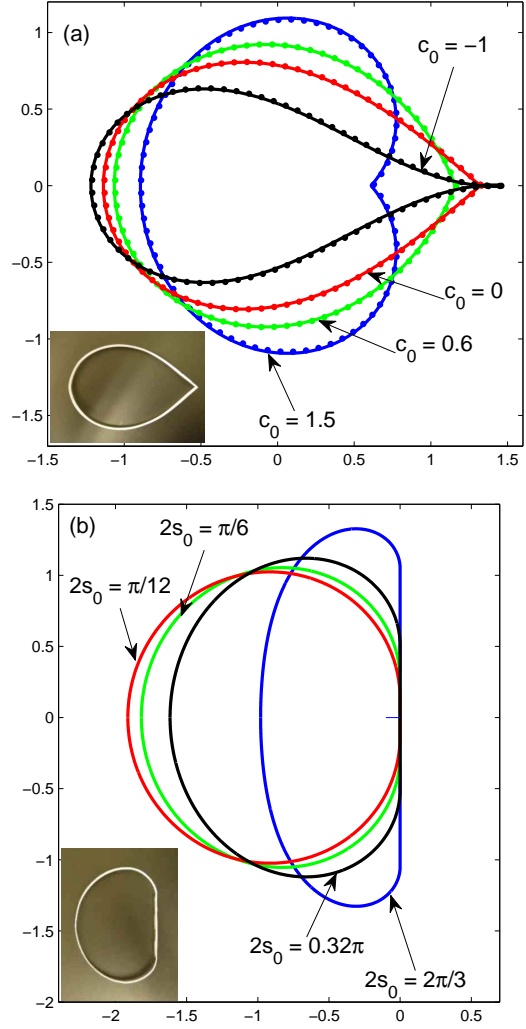}
\caption{\label{fig:ringsnopress} (a) Rings with an infinitely weak point. The spontaneous curvature $c_0$ is constant along each ring. Curvature of the black ring: $c_0=-1.$, red: $0.0$, green: $0.6$ and blue: $1.5.$. The dots represent the result of a tethered chain energy minimization simulation with $N=601$ points (only a fraction of the points is displayed). Inset: photograph of a thin ribbon of paper attached at its ends by flexible tape.  (b) Rings of zero spontaneous curvature ($c_0=0$) with an infinitely strengthened segment. The circumference of the ring measures $2\pi$ and the length of the strengthened segment is $2s_0$. Blue curve: $2s_0= 2\pi /3$, green $\pi /6$, red $ \pi / 12$ ,black $ 0.32 \pi$. Inset: photograph of a paper ring with a strengthened segment equal to  $16\%$ of the total ring length. 
(Color available online)}
\end{figure}

In the absence of hydrostatic pressure, the shape of a ring with an infinitely weak link (i.e. a rod of spontaneous curvature $c_0$ hinged at its ends) can be determined exactly, and the solution can be expressed in closed form in terms of elliptic integrals. A simple desktop approximation is a piece of writing paper, gently folded back and attached to itself with flexible tape (see inset of Fig.~\ref{fig:ringsnopress}(a) and Fig.~\ref{fig:explX}(a)). These solutions are closely related to the elastica solutions discussed in Ref.~\cite{Love02}. However, in our case the boundary conditions are fixed and the control parameter is the spontaneous curvature rather than the force applied on a rod.

We set the origin of the arc-length $s=0$ at the infinitely weak point and orient the ring so that it has bilateral symmetry across the $x$-axis (see Fig.~\ref{fig:explX}).
At the infinitely weak point the bending energy, and hence the torque, is zero. The ring curvature at $s=0^+$ (and $s=2\pi^-$) thus has to equal the spontaneous curvature $c_0$. For the piece of initially flat writing paper mentioned above, we have $c_0=0$ at this point. The continuity of the tangent angle at $s=\pi$ and the bilateral symmetry of the ring dictate that the ring has to cross the $x$ axis at right angles at $s=\pi$.

These considerations translate to the following boundary conditions:
\begin{eqnarray}\label{bccut2}
\frac{\dd\theta(s)}{\dd s} \bigg|_{s=0^+}&=&c_0\nonumber\\
\theta(\pi)&=&3\pi/2.
\end{eqnarray}
Last, the bilateral symmetry and orientation that we have chosen for the ring implies $y(0) = y(\pi)$, so the tangent angle $\theta(s)$ needs to satisfy:
\begin{eqnarray}\label{bccut3}
y(\pi)-y(0)=\int\limits_0^{\pi}\sin \theta(s)\dd s = 0.
\end{eqnarray}
To determine the tangent angle subject to the boundary conditions of Eq.~\eqref{bccut2}, we need to functionally minimize the ring bending energy, and implement the constraint of Eq.~\eqref{bccut3} with a Lagrange multiplier.

For $c_0<1$ (preferred radius of curvature $c_0^{-1}$ greater than the actual one $r_c$), after integrating the differential equations that are derived from the functional minimization of Eq.~\eqref{bendingenergy1}, the function $\theta(s)$ is found to be:
\begin{equation}\label{bccut4}
\theta(s)=\frac{3\pi}{2}-2\; \mathrm{am}\left[k\lambda(1-s/\pi),1/ k\right],
\end{equation}
where $\mathrm{am}\left[x,k\right]$ is the Jacobi amplitude \cite{Abramowitz72}.
The constants $k$ and $\lambda$, which enter the equations as an integrating constant and a Lagrange multiplier respectively, are determined by the equations:
\begin{eqnarray}\label{bccut5}
E\left[\cos^{-1} \left(\frac{\pi c_0}{2k\lambda}\right) ,k\right]&=&\lambda/2\nonumber\\
F\left[\cos^{-1} \left(\frac{\pi c_0}{2k\lambda}\right),k\right]&=&\lambda,
\end{eqnarray}
where $E\left[x,k\right]$ is the incomplete elliptic integral of the second kind and $F\left[ x,k\right]$ is the incomplete elliptic integral of the first kind \cite{Abramowitz72}. For $c_0\rightarrow 1$ we retrieve, as expected, a perfect circle, as the hinge at $s=0$ has no effect. (To recover results for $r_c\ne 1$, simply let $c_0\rightarrow c_0 r_c$ in the above formulas.)

The above equations also hold for $c_0<0$, when the spontaneous curvature is negative (preferred radius of curvature points outwards, which can be achieved by inverting a positive spontaneous curvature ring). Note that for a spontaneous curvature below a certain threshold (approximately equal to $c_0=-1$), the solutions of Eq.~\ref{bccut4} will correspond to self-intersecting rings.

For a ring with zero spontaneous curvature, Eqs.~\eqref{bccut4} and \eqref{bccut5} simplify to:
\begin{eqnarray}
\theta(s)&=&\frac{3\pi}{2}-2 \; \mathrm{am}\left[k\;\mathrm{K}(k)(1-s/\pi), 1/k\right] \\
\mathrm{E}(k)& = &\frac{\mathrm{K}(k)}{2},
\end{eqnarray}
where $K(k)$ is the complete elliptic integral of the first kind. For a zero spontaneous curvature ring, the angle of the cusp at the weak point reads $\psi_0 = 2\left[\pi-\theta(0^+)\right]=4\sin^{-1}k-\pi$, which equals approximately $\psi_0\simeq 0.4523 \pi$. (See Fig.~\ref{fig:explX}(a) for the definition of $\psi_0$.)

For $c_0>1$ (preferred radius of curvature $c_0^{-1}$ less than the actual one), Eq.\eqref{bccut4} and Eq.\eqref{bccut5} read:
 \begin{equation}\label{bccut4b}
\theta(s)=\frac{\pi}{2}- 2 \; \mathrm{am}\left[k \lambda (1 - s/\pi)-\mathrm{K}(k) , 1/k\right];
\end{equation}
and
\begin{eqnarray}\label{bccut5b}
\frac{\lambda }{2k}-k\lambda &=&\text{E}\left[ \text{am}\boldsymbol(\text{K}\left(1/k \right)-k \lambda ,1/k \boldsymbol),1/k \right]- \text{E}\left(1/k \right)\nonumber\\
\frac{\pi  c}{2 \sqrt{k^2-1} \lambda }&=&\text{nd}\left(\lambda  k,1/k\right).
\end{eqnarray}
where $\text{nd}(x,k)\equiv \frac{1}{\sqrt{1-k^2 \text{sn} ^2(x,k )}}$ is the Jacobi nd function \cite{Abramowitz72}.

In Fig.~\eqref{fig:ringsnopress}(a) we plot ring profiles, obtained by evaluating the analytic expressions Eq.\eqref{bccut4} and Eq.\eqref{bccut4b} assuming an infinitely weak point and a variety of spontaneous curvatures. As the curvature approaches $c_0 = 1$ the rings tend to become perfectly circular. The curvatures of the black, red, green and blue ring are respectively $c_0=-1.$ (inverted ring), $0.0$, $0.6$ and $1.5$. The dots represent the result of a tethered chain simulation with $N=601$ points (along the lines of Sec.~\ref{sec:Collapse} and Ref.~\cite{Leibler87}; for clarity only a fraction of the points is displayed). Note that these ring profiles were obtained assuming self-intersections are possible. However, the minimum energy shapes discussed here do not require self-intersections, except for $c_0\lesssim -1$ or very large positive $c_0$. In these cases the ring edges near $s=0$ will have $y(s)=0$ until an arc-length $s_c$, and the overall ring shape can then be determined by adjusting the boundary conditions at the origin and rescaling the ring (see inset of Fig.~\ref{fig:explX} for an illustration).

The zero spontaneous curvature case is the simple case of a flexible rod, straight in the absence of torques, and hinged at its ends. For comparison with the case $c_0=0$, the inset of Fig.~\eqref{fig:ringsnopress}(a) is a picture of ring constructed from a ribbon of paper (zero spontaneous curvature case) and attached at the ends with a thin segment of adhesive tape.

\subsection{\label{sec:strong_e}Ring with infinitely strong sector}

As in Sec.~\ref{sec:weak_e}, in this section we find an exact solution to the equations derived from functional minimization of Eq.~\eqref{bendingenergy1} for rings of zero spontaneous curvature, $c_0=0$. However, this time we consider rings with an infinitely strengthened segment ($\mathrm{\kappa}(s)\rightarrow\infty$) of length $2s_0$, over which the curvature is constrained to be exactly zero (see Fig.~\ref{fig:explX}(b)).

Upon orienting the ring so that the strengthened segment  coincides with the $y$-axis and the $x$-axis intersects the ring in the middle (see Fig.~\ref{fig:explX}(b)), we immediately see that continuity of the tangent angle $\theta(s)$ imposes the following boundary conditions:
\begin{eqnarray}\label{bcstrong2}
\theta(s_0)&=&\pi/2 \nonumber\\
\theta(\pi)&=&3\pi/2.
\end{eqnarray}
The continuity of $y(s)$ and bilateral symmetry of the ring now implies:
\begin{equation}\label{contystrong}
y(s_0)-y(\pi)=s_0\Rightarrow \int\limits_{s_0}^{\pi}\sin{\theta(s)}\dd s = -s_0.
\end{equation}
Note that $y(\pi) = 0$ for the configuration of Fig.~\ref{fig:explX}(b).
The $\theta(s)$ angle that minimizes the energy functional now reads:
\begin{equation}\label{exactstrong}
\theta(s)=\frac{\pi} {2}+ 2\;\mathrm{am}\left[\frac{c_1}{2}(s - s_0),k\right],
\end{equation}
where $c_1$, the curvature at $s_0$, and the integration constant $k$ are determined by Eq.~ \eqref{bcstrong2} and the continuity condition of Eq.~\eqref{contystrong}. We can recast these two equations as:
\begin{equation}\label{bcstrong3}
c_1 = \frac{2 K(k)}{\pi-s_0}
\end{equation}
and
\begin{equation}\label{contystrong2}
\frac{K(k)}{2(1-s_0/\pi)}=\frac{1}{k^2}\left[K(k)-E(k)\right].
\end{equation}
Eq.~\eqref{contystrong2} can be solved numerically to determine $k$ and subsequently $c_1$ as functions of $s_0$.
For $s_0=0$, the case of a ring with infinitesimally short strong sector, the solution of Eq.~\eqref{contystrong2} is $k=0$, so $c_1=1$ and $\theta(s)=\pi/2+s$, the tangent angle of an undeformed circular ring.

In Fig.~\eqref{fig:ringsnopress}(b) we plot ring profiles, obtained by evaluating the analytic expressions Eq.\eqref{exactstrong}  assuming an infinitely strong sector of varying lengths and $c_0 = 0$. As the length of the strong segment approaches $s_0 = 0$ the ring tends to become perfectly circular. The length of the strong segment of the red, green, black and blue ring is respectively $2s_0=\pi/12$, $\pi/6$, $0.32$ and $2\pi/3$. 

\subsection{\label{sec:small_nonuni}Ring with small non-uniformities}

The boundary conditions and symmetry considerations of Sec.~\ref{sec:weak_e} and \ref{sec:strong_e} cannot be used to determine the shape of a ring at zero pressure with an arbitrary non constant bending modulus. We now assume that $\kappa(s)$ is smoothly varying, and then consider small deviations from uniformity.
In this case, after functional minimization of the bending energy with the closure constraints (see App.~\ref{appeqderiv}), we get:
\begin{equation}\label{small_nonunieq}
\frac{1}{c(s)}\left[\frac{\left[\kappa(s) (c(s)-c_0)  \right]'' }{c(s)}    \right]'  + \left[\kappa(s) (c(s)-c_0)  \right]' =0
\end{equation} 
where the prime stands for differentiation with respect to the arc-length.
If $\kappa_0$ is the average value of $\kappa(s)$ and we assume that both $\dk\equiv \kappa(s) -\kappa_0$ and $\Delta c(s)$ are small, we can expand Eq.~\eqref{small_nonunieq} to linear order in $\dk$ and $\Delta c$ to get:
\begin{equation}\label{small_nonunieq_linear}
\kappa_0 (\Delta c'' + \Delta c) = (c_0 - 1)(\dk''+ \dk) + b
\end{equation} 
where $b$ is an integration constant.

From the solution of Eq.~\eqref{small_nonunieq_linear} one can derive the tangent angle that determines the equilibrium zero pressure shape of the ring  (see App.~\ref{appthetanop}):
\begin{equation}\label{non_uni_thetas}
\theta(s) = s+ \alpha_1 \cos s + \alpha_2 \sin s +\frac{(c_0-1)}{\kappa_0} \int\limits_{-\pi}^s \dd\xi \dk(\xi) + \theta_0.
\end{equation}
$\alpha_1$, $\alpha_2$ are constants from the homogeneous part of the solution, and $\theta_0$ is an integration constant. The periodicity constraint Eq.~\eqref{constraint1} fixes the values of $\alpha_1$, $\alpha_2$ to (see App.~\ref{appthetanop}): 
\begin{equation}\label{alpha1}
\alpha_1 = \frac{c_0-1}{\pi\ka_0}\int\limits_{-\pi}^{\pi}\dd s\;  \dk(s)\; \sin s
\end{equation}
and
\begin{equation}\label{alpha2}
\alpha_2 = -\frac{c_0-1}{\pi\ka_0}\int\limits_{-\pi}^{\pi}\dd s\; \dk(s)\; \cos s. 
\end{equation}
The constant $\theta_0$ can be determined if we consider that the maximum value of $|c-c_0|$ should occur at the minimum $\kappa(s)$. 
$\dk$ always appears multiplied by $c_0-1$ in this small deformation regime. Thus, if $c_0$ equals $1$ (the curvature of the circular undeformed ring), then $\theta(s) = s+\theta_0$, which is the tangent angle of an undeformed circular ring. Hence, if the spontaneous curvature $c_0$ equals the undeformed ring curvature, then at zero pressure the ring remains circular, \textit{even at the presence of inhomogeneities}. Inverting the sign of $c_0-1$ (by, e.g., changing the spontaneous curvature from $0$ to $2$) is equivalent to inverting the sign of $\dk$, as the small curvature segments exchange place with the large curvature segments. 

\begin{figure}
\includegraphics[scale=0.85]{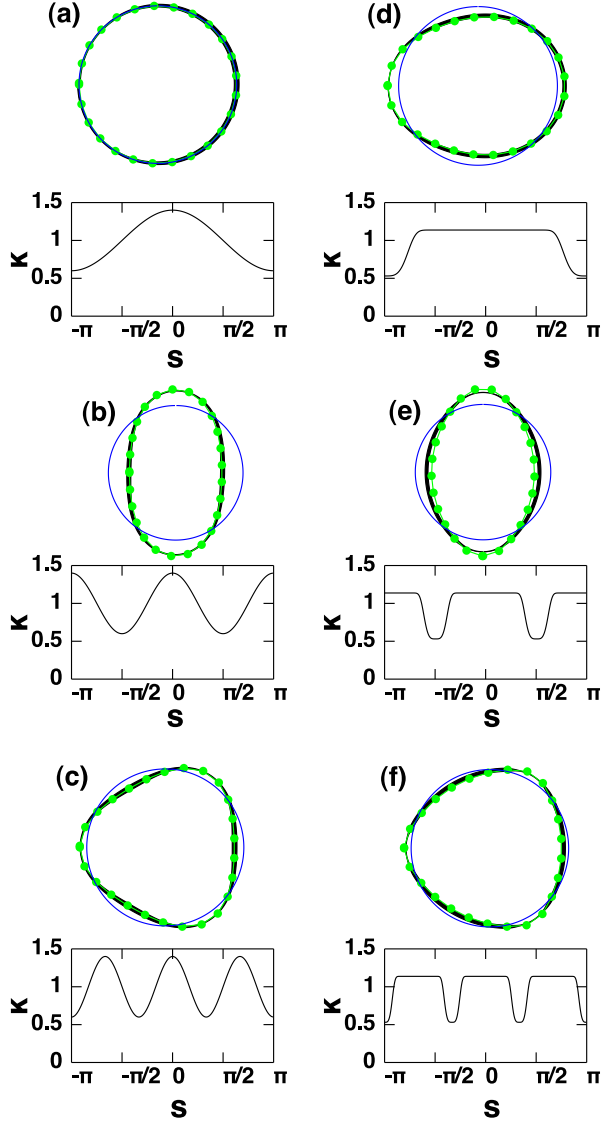}
\caption{\label{fig:invertcurv} Zero pressure equilibrium configuration of rings with $c_0\ne 1/r_c=1$ and non constant bending modulus. In all cases we consider ``inverted" rings with $c_0=-1$. The blue line represents a circle of the same circumference, and the distorted shapes allow optimal release of bending energy. All rings remain circles unaffected by the variation in $\kappa(s)$ when $c_0=1$. The dotted green line is the result of tethered chain simulations. To illustrate the spatial variation in $\ka(s)$ the thickness of the rings is drawn is proportional to $(0.5+\delta\kappa(s))/15$. 
(Color available online)  }
\end{figure}

In Fig.~\ref{fig:invertcurv} we plot six cases of rings with $c_0\ne 1$ and nonuniform bending modulus. We focus on the particularly interesting case $c_0=-1$, which corresponds physically to inverting a circular ring (initially with $c_0=1$) cut, e.g., from a yogurt cup. The ring as cut ($c_0=1$) is insensitive to variations in the dimensions along the symmetry axis ($\dk(s)\ne 0$), but these become quite pronounced when the ring is inverted ($c_0=-1$). In all rings of Fig.~\ref{fig:invertcurv} $c_0= -1$, and $\ka(s)$ with $\kappa_0=1$ is shown below each shape. The dotted green line is the result of tethered chain simulations (see Sec.~\ref{sec:Collapse}), the solid black line (the thickness of which corresponds to the magnitude of the bending modulus) is the analytical result, and the blue line is a circular ring of the same circumference, shown for comparison. All rings coincide with this circular blue ring for $c_0=1$. Note that for a sinusoidally modulating $\ka(s)$ with period $2\pi$, the $p=0$ configuration of the ring is circular, as seen in Fig.~\ref{fig:invertcurv}(a). The same is not true for all rings with $\kappa(s)$ periodicity equal to $2\pi$, e.g. the ring in Fig.~\ref{fig:invertcurv}(d).

As seen from the bending modulus plots, the plotted rings are far from the small $\dk$ regime, when we expect the analytic result to be accurate. Nevertheless, we observe excellent agreement between the analytic results and the simulations, so, provided that $\dk$ does not change too abruptly,   Eq.~\eqref{non_uni_thetas} gives a reasonable prediction for the unpressurized ring shape.

\section{\label{sec:Criticalpress}Expansion in Fourier modes and critical pressure of nonuniform rings}

In this section we discuss ``unfrustrated'' pressurized rings with spontaneous curvature $c_0$ equal to the undeformed ring curvature $1/r_c$ ($c_0=1$ in the units of Sec.~\ref{sec:Spontcurv}) and an arc-length dependent bending modulus. We determine the critical buckling pressure $p_c$ for various bending modulus profiles and explore the properties of $\kappa(s)$ that control the ring stability.

\subsection{Quadratic expansion}

As before, we describe the shape of a ring with arbitrary $\kappa(s)$ by the tangent angle of a circular reference ring, $\theta_0 (s) = \mathrm{const} +s$ plus a function  $\dth(s)$ which describes the deviation from the circular shape:
\begin{equation}\label{dtheta}
\theta (s) = \theta_0 +s + \delta \theta (s).
\end{equation}
The constant $\theta_0$ can be chosen arbitrarily and just corresponds to an overall rotation of the ring. All lengths are henceforth measured in units of the ring radius, so the total ring circumference equals $2\pi$.

The function $\dth(s)$ can be expanded in Fourier modes:
\begin{equation}\label{dthetafourrier}
\delta \theta (s) = \sum\limits_{n =  - \infty }^\infty  {\delta \theta _n e^{ - ins} }.
\end{equation}
Note that since $\delta \theta (s) $ is real, the complex Fourier coefficients satisfy
\begin{equation}\label{realityconstraint}
\delta \theta _n =\delta \theta _{-n}^*.
 \end{equation}
The zero mode just corresponds to an overall rotation of the ring, so we set it equal to zero. (Upon choosing the constant in Eq.~\eqref{dtheta} to be $\theta_0=\pi/2$, we would get  $\theta(-\pi)=-\pi/2$, consistent with the parameterization in Sec.~\ref{sec:Spontcurv}.)
With this parameterization the inextensibility constraint is satisfied automatically.
However, the modes $\delta \theta _n$ are constrained by the periodicity conditions of Eq.~\eqref{constraint1}, which translates to a complicated nonlinear condition on the Fourier modes:
\begin{equation}\label{constraint3}
 \sum_{N=1}^{\infty}\frac{i^N}{N!}\sum\limits_{n_1,n_2,\dots,n_{N}}\delta\theta_{n_1}\delta\theta_{n_2}\dots\delta\theta_{n_{N}}\delta_{1,\sum_{j=1}^N n_j}=0.
\end{equation}
Here the summation of $n_1\dots,n_{N}$ extends from $-\infty$ to $\infty$, and $\delta_{k,l}$ is the Kroenecker delta.

In what follows we discuss rings whose unpressurized state is circular, i.e. the spontaneous curvature is constant and equal to the inverse of the ring radius. As discussed in Sec.~\ref{sec:Spontcurv}, the ground state \textit{remains} circular when $c_0=1$, in the absence of an external pressure. A small deviation from the circular, natural shape of the ring, as for example in the postbuckling regime for the ring when $p \rightarrow p_c^+$, thus corresponds to small Fourier coefficients $\dth_n$.

To second order in $\delta\theta$, the constraint Eq.~\eqref{constraint3} reads:
\begin{equation}\label{constraint}
\delta\theta_1+\frac{i}{2}\sum_{n=-\infty}^{\infty}\delta\theta_n\delta\theta_{1-n}=0.
\end{equation}
Note that $\dth_1$ is the only mode that appears at linear order, and for small deviations $\dth(s)$ we expect that we can simply set $\dth_1=\dth_{-1}\equiv 0$ \cite{Panyukov01}. More generally  $\dth_1$ is actually quadratic in the remaining modes, due to Eq.~\eqref{constraint}.

In the general case where the the bending modulus $\kappa(s)$ is an arc-length dependent constant and $c_0=1$, the bending energy in Eq.~\eqref{bendingenergy1} reads:
\begin{equation}\label{bendingenergy2}
E_{\mrben}[\theta (s)] = \mathrm{const} +\frac{1}{2}\int\limits_{ - \pi }^\pi ds \kappa(s) \left( \frac{\dd \delta\theta (s)} {\dd s} \right) ^2.
\end{equation}
The spontaneous curvature and the constant part of the derivative of $\theta(s)$ cancel when $c_0=r_c^{-1}$, so the bending energy is quadratic in $\dth$ with no linear terms.

The nonuniform bending modulus $\kappa(s)$ can be expanded in Fourier modes as:
\begin{equation}\label{dkappafourrier}
\kappa(s) = \sum\limits_{n =  - \infty }^\infty  {\kappa _n e^{  ins} }.
\end{equation}
The average bending modulus of the ring will be denoted as: $\kappa_0  = \frac{1}{2\pi}\int_{ - \pi }^\pi  \kappa(s) ds$.

Upon expressing the bending energy in terms of $\ka$ and $\dth$ modes we get:
\begin{equation}\label{ebending2}
E_{\mrben}[\{\delta\theta_n \}] = \pi \sum_{|m|,|l|\ne 0,1} m\:l\:\kappa_{l-m}\delta\theta_m^* \delta\theta_l.
\end{equation}
As already discussed, the mode $\dth_0$ is explicitly excluded because it corresponds to an overall rotation of the ring. In fact, allowing $l,m=0$ would not change the value of the sum since all the terms containing $\dth_0$ are multiplied by zero. In addition, we have excluded $|m|,|l| =  1$, since the $\dth_{\pm 1}$ mode is suppressed to lowest order in the nonlinearity due to the constraint of Eq.~\eqref{constraint}. 

Upon suppressing the constant in Eq.~\eqref{bendingenergy2} , Eq.~\eqref{ebending2} can be expressed in matrix form:
\begin{equation}\label{}
E_{\mrben}[\{\dth_n \}] = \boldsymbol{\dth}^{\dagger} \hat{E}_{\mrben} \boldsymbol{\dth}
\end{equation}
where $\boldsymbol{\delta\theta}$ is the infinite dimensional column vector of $\delta\theta_n$ coefficients, and the matrix elements of $\hat{E}_{\mrben}$ are:
\begin{equation}\label{bendmels}
\hat{E}_{\mrben_{m n}} = \pi\; m\; n\; \ka_{n-m}.
\end{equation}

 As in uniform rings, we expect that when $|m|\gg 1$, $\dth_m$ modes will contribute very little to the buckling behavior and overall shape everywhere below (and slightly above) the buckling transition. This observation allows us to require $m\le M$ and truncate the infinite dimensional $\boldsymbol{\dth}$ to a $2(M-1)$-component vector, so that
\begin{equation}\label{dtheta_vec}
\bm{\dth} = \left[ \begin{array}{c} \dth_{-M}  \\ \dth_{-M+1}  \\ \dots  \\ \dth_{-3} \\ \dth_{-2} \\ \dth_{2}\\ \dots \\ \dth_{M}  \end{array} \right].
\end{equation}
As discussed above, the amplitudes $\dth_0$ and $\dth_{\pm 1}$ are set to zero. We shall similarly truncate the bending energy expansion of Eq.~\eqref{dkappafourrier} to include only Fourier modes up to $|n| \le M$. This truncation if a good approximation for smoothly varying rings, i.e. for rings with $|\ka_m|\ll \ka_0$ for $m\gtrsim M$. 

Note that, although the $\dth_n$ modes are complex, the condition of Eq.~\eqref{realityconstraint} reduces the complex number of degrees of freedom represented by the vector $\bm{\dth}$ by half, from $4(M-1)$ to $2(M-1)$.

The (Hermitian) truncated matrix $\hat{E}_{\mrben}$ looks like:
\begin{widetext}
\begin{equation}\label{toeplitz}
\hat{E}_{\mrben} = 
\left[ \begin{array}{c c c c c c c c c c}
M^2 \kappa_0, & M(M-1)\ka_1, &.. & 2M\ka_{M-2},   & 0 &0& \dots &  ..&0 \\
{}      & (M-1)^2 \kappa_0, &.. &2(M-1)\ka_{M-3},& 0 &0 & \dots& ..&0 \\
{}      & {}          &.. & \dots             & \dots & \dots & \dots & ..& \dots\\
{}      & {}          &{}    & 9 \kappa_0,             &6\ka_1, &-6\ka_5, & -9\ka_6,&..& \dots\\
{}      & {}          & {}   & {}                &4 \kappa_0, & -4\ka_4,&-6\ka_5, &..&0 \\
{}      & {}          &         & {}      & {}        & 4 \kappa_0,  &6\ka_1, & .. &2M\ka_{M-2}\\
{}      & c.c.          &      &        &              &       & 9 \kappa_0, & .. &3M\ka_{M-3}\\
        &                 &      &         &             &       &                        &   .. &\dots     \\
      &                  &       &         &            &       &                         &          &M^2 \kappa_0
\end{array} \right]
\end{equation}
\end{widetext}

 The pressure term in the total energy ($p$ multiplies the area, given by Eq.~\eqref{area}) depends on $\dth(s)$ in a more complicated, non-local fashion. To express $E_{\mrpr}$ in terms of the $\dth_n$ modes, we need to perform a Taylor expansion of the area in Eq.~\eqref{area} in small $\dth(s)$, substitute with the Fourier expansion of $\dth(s)$ and perform the double integral in the arc-length. At first sight in this expansion, the $\dth_1$ mode seems to appear in linear order, and therefore cannot be disregarded. However, the $\dth_1$ and $\dth_{-1}$ modes that appear in linear order cancel with sums of higher order terms as dictated by Eq.~\eqref{constraint}.

A more convenient approach to performing the expansion in $\dth_n$ modes is to write Eq.~\eqref{area} for the area $A$ of the ring as:
\begin{equation}\label{}
A = \frac{1}{2}\mathrm{Im} \int\limits_{-\pi}^{\pi} ds \int\limits_{-\pi}^{s} d\xi e^{i (s-\xi)}e^{i(\dth(s)-\dth{\xi})}.
\end{equation}
If we expand the quantity $e^{i\dth(s)}$ in Fourier modes as:
\begin{equation}\label{}
e^{i\dth(s)} = \sum_n f_n e^{-ins}
\end{equation}
we can immediately see that the periodicity constraint exactly translates to setting $f_1=0$. Notice that, in general $f_n\ne f_{-n}^*$ in this representation.
With the help of the $f_n$ modes, we can express the area exactly as:
\begin{equation}\label{}
A = \pi \left(|f_0|^2 - \sum_{n\ge 2} |f_n|^2 \frac{2}{n^2-1}\right).
\end{equation}
We can then straightforwardly expand $f_n$ in $\dth_m$ modes and collect terms of equal order. We therefore see that the excess area $\Delta A = A - \pi$ is given by:
\begin{equation}\label{}
\Delta A = - 2\pi \left(|\dth_1|^2 + \sum_{m\ge 2} |\dth_m|^2 \frac{m^2}{m^2-1}\right) + O(\dth^3).
\end{equation}
Because of Eq.~\eqref{constraint}, $|\dth_1|^2$ is of higher order and can be disregarded\footnote{Note that shift in area is always negative, confirming that a perfect circle is indeed the maximum area for a given perimeter.}. We see that the area, up to quadratic order, is diagonal in the $\dth_m$ modes. Upon defining an area difference matrix as $\Delta \hat{A}$ such that $\Delta \hat{A}_{mn} = -\delta_{mn} \pi \frac{m^2}{m^2-1}$, the total elastic energy matrix of the ring reads:
\begin{equation}\label{totalmatrixene}
\hat{E} = \hat{E}_{\mrben}+p\Delta \hat{A}.
\end{equation}

At zero pressure the energy is always positive irrespective of the ring deformation, so the matrix $ \hat{E}_{\mrben}$ is positive definite. The circular shape of the ring which corresponds to the trivial configuration $\dth(s)=0$, is a stable energetic minimum. As the pressure increases, at a critical pressure $p_c$ the lowest eigenvalue of $\hat{E}$ (or eigenvalues if the lowest energy eigenstate is degenerate) becomes zero, and with further increase of the pressure, negative. The circular shape becomes unstable at a critical pressure and the ring begins to buckle. Determining the critical pressure for the onset of buckling is therefore equivalent to determining the pressure at which the matrix $\hat{E}$ first acquires a negative eigenvalue.

This requirement can be translated to something simpler if we consider the diagonal matrix whose elements are $K_{mn} = \delta_{mn} \left(\frac{m^2-1}{\pi m^2}\right)^{1/2}$. This matrix is positive definite, and if $\hat{E}$ is also positive definite then the product $\hat{K}\hat{E}\hat{K}$ should also be positive definite.
But it is easy to see that $\hat{K}\hat{E}\hat{K} = \hat{K}\hat{E}_{ben}\hat{K} - p\hat{I}$, where $\hat{I}$ is the identity matrix. Thus determining the critical pressure reduces to the easier problem of finding the smallest eigenvalue of the Hermitian matrix $\hat{B}=\hat{K}\hat{E}_{\mrben}\hat{K}$. We can now proceed to diagonalize $\hat{B}$ and determine the critical pressure $p_c$ for rings with different $\kappa(s)$. 

\subsection{Bending modulus parameterization}

The full space of $\kappa(s)$ functions that we could explore is infinite dimensional. To simplify matters we identify the set of parameters that we expect to be qualitatively important in explaining the trends in the value of $p_c$ for a variety of inhomogeneous rings.

We shall compare rings with the same average bending modulus $\kappa_0$. Assume for simplicity that the ring is carved from a substance with uniform bulk material properties. The bending modulus for in-plane deformations of thin rings of a rectangular cross-section is then proportional to $\ka(s)\sim h_r(s)^3h_z(s)$, where $h_r(s)$ is the thickness of the ring along its radius and $h_z(s)$ the thickness along the perpendicular to the plane dimension \cite{Landau70}. The linear mass density is proportional to $\dd m/\dd s\sim h_r(s) h_z(s)$. Assuming that the $s$-variation in the bending modulus comes through variation in the thickness $h_z(s)$, the average bending modulus is proportional to the total mass of the ring. Similarly, keeping $h_z(s)$ constant and varying $h_r(s)$, $\ka_0$ is proportional to $\ka_0\sim\int_{-\pi}^{\pi}\dd s\large(\frac{dm}{ds}\large)^2$. Hence, for the same total amount of material (average $h_r(s)$), the maximum average bending modulus can be achieved for rings with constant thickness. Alternatively, more mass is required for rings with the same average bending modulus if we increase the amplitude of the $\kappa(s)$ modulations. 

In this subsection we focus primarily on rings with \textit{periodic} modulations in $\kappa(s)$ (rings with quenched random variations in the bending modulus will be discussed later). For $n$-fold periodic rings, $\kappa(s+ 2\pi / n) = \kappa(s)$. As an immediate consequence of the $2\pi/n$ periodicity, the only modes that do not vanish in the expansion of Eq.~\eqref{dkappafourrier} are integral multiples of $n$. We will consider functions $\kappa(s)$ with only one (global) minimum within each period $t_s = [-\pi/n, \pi/n )$ and reflection symmetry of the ring across the line that passes through the minimum and the center of the ring. Without loss of generality we can choose the arc-length origin of the ring so that $\kappa(0) = \min\kappa(s)$. It follows that $\kappa(\pi/n)=\kappa(-\pi/n)=\max\kappa(s)$.  For simplicity we define $w_{\min} \equiv \min{\kappa(s)}$ and  $w_{\max} \equiv \max{\kappa(s)}$. We now reparametrize the function $\kappa(s)$ for $s\in  t_s$ in terms of a function $G(x)$ that parametrizes the shape of the defect:
\begin{equation}\label{Gofx}
G(s/\alpha) = \frac{w_{\max}-\kappa(s)} {w_{\max}-w_{\min}}.
\end{equation}
Here, $\alpha$ is the width of a "defect", which we take to be a weakened ring segment similar to the $\kappa(s)$ profiles in Figs.~\ref{fig:invertcurv}(a)-\ref{fig:invertcurv}(f).
By definition $G(0) = 1$ and $G(\pi/n\alpha) =G(-\pi/n\alpha) = 0$. Upon defining  $G(|s|>\pi/n\alpha) \equiv 0$ we can extend the definition of $G(x)$ to the whole real axis. After averaging Eq.~\eqref{Gofx} over the interval $t_s$ we find: 
\begin{equation}\label{Gofxave}
n \alpha q = 2\pi \frac{w_{\max}-\kappa_0} {w_{\max}-w_{\min}},
\end{equation} 
where a normalization is provided by 
\begin{equation}
q \equiv \int_{-\infty}^{\infty} G(x)\dd x. 
\end{equation}
Another quantity that can be used to quantify the shape of the $\kappa(s)$ function  is its standard deviation relative to $\kappa_0$, averaged around the ring:
\begin{equation}\label{Gofxsigma}
\sigma = \sqrt{<(\kappa(s)-\kappa_0)^2>} = \sqrt{2\sum_{m=1}^{\infty}{ |\ka_m|^2}}.
\end{equation} 
Upon using Eq.~\eqref{Gofx}, we can recast Eq.~\eqref{Gofxsigma} as:
\begin{equation}\label{Gofxsigma2} 
\sigma = (w_{\max}-w_{\min})\frac{n\alpha}{2\pi}\sqrt{\frac{2\pi}{n\alpha} q_{\sigma}-q^2},
\end{equation}
where $q_{\sigma}  \equiv \int_{-\infty}^{\infty} G(x)^2 \dd x$. 

A list of three different functions $G(x)$ describing weakened sectors and their $q$ and $q_{\sigma}$ values is given in Table~\ref{tab:table1}. As the $m$ exponent of the gaussian type function (second row of the table) increases, the function $G(x)$ changes from a smooth, slowly varying function to a step like one. Increasing $m$ in the second row thus interpolates between the smooth variation in the first row and the abrupt one in the third.

\begin{table}
\caption{\label{tab:table1} $G(x)$ functions describing weakened sectors and their $q$ and $q_{\sigma}$ values.}
\begin{ruledtabular}
\begin{tabular}{ | c  | c | c |}
\hline
$G(x)$ & q  & $q_{\sigma}$  \\ \hline\hline
$\frac{\cos(\pi s)+1}{2}, \; t_s=[-1,1)$ & $1$ & 3/4  \\  \hline
$\exp(-(2x)^m),\; m$ even & $\Gamma(\frac{m+1}{m})$ & $\frac{1}{2^{1/m}}\Gamma(\frac{m+1}{m}) $ \\ \hline
$\Theta(1/2-|x|)$ & 1 & 1  \\ 
\hline
\end{tabular}
\end{ruledtabular}
\end{table}

\subsection{Stability of rings with a periodic bending rigidity}\label{stabilityofthering}

\begin{figure*}
\includegraphics[scale=0.75]{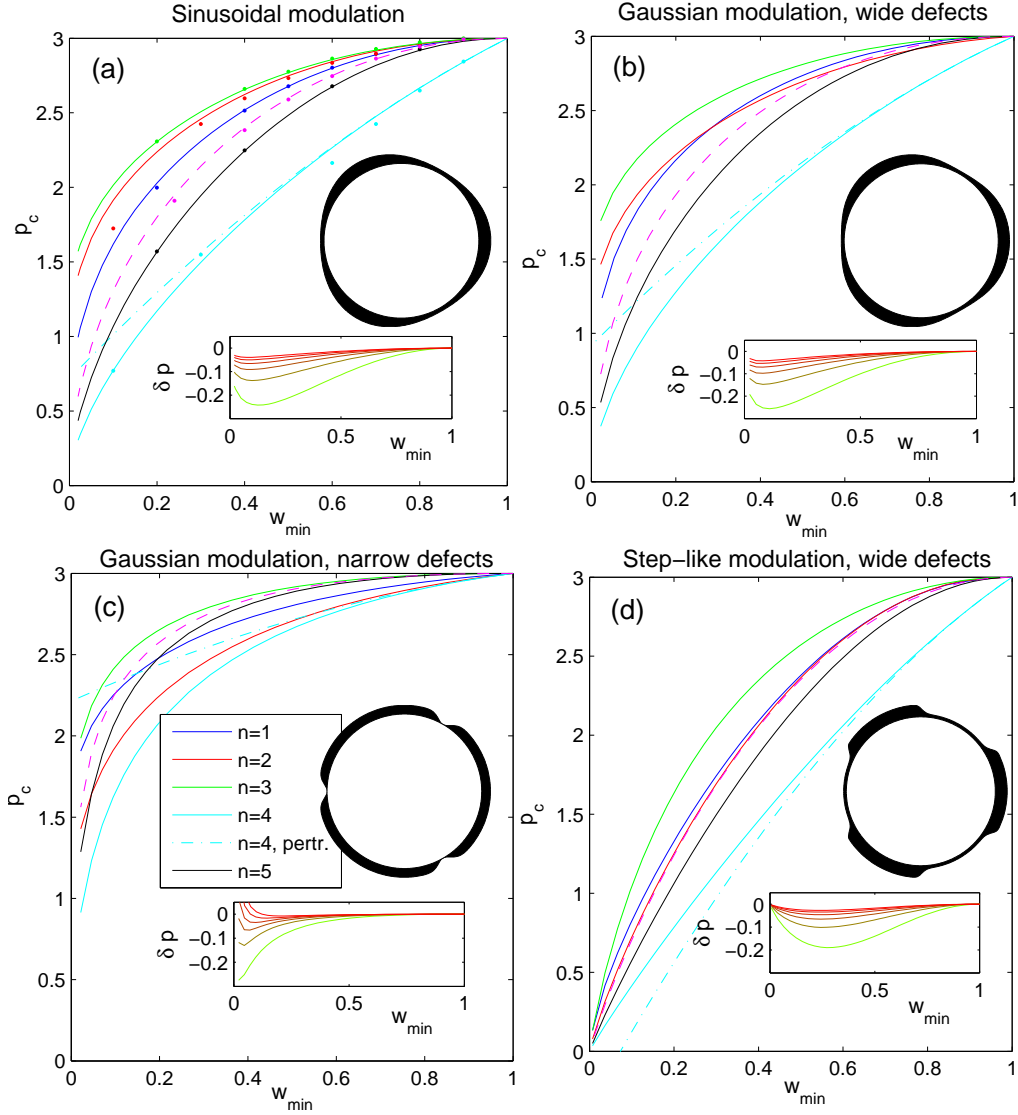}
\caption{\label{fig:minwidth}  Dependence of the critical pressure on defect amplitude: we plot the $w_{\min}$ dependence of critical pressure for various ring profiles. Here $w_{\min}$ is measured in units of $\ka_0$, the average bending rigidity of the ring. (a) $G(x)=\frac{\cos(\pi s)+1}{2}$, (b) $G(x) = \exp(-(2x)^2)$,  $\alpha=\pi/2$, (c) $G(x) = \exp(-(2x)^2)$,  $\alpha=\pi/8$, (d) $G(x) = \exp(-(2x)^{12})$,  $\alpha=\pi/2$. Although 3-fold symmetric rings are shown, the data reflects calculations for $n=1-5$. Blue line: $n=1$, red line: $n=2$, green line: $n=3$, cyan line: $n=4$, black line: $n=5$ (Color available online). The dashed cyan line is the perturbation theory prediction for $n=4$. Note that the curves are not monotonic with $n$ near $n=4$. As $n$ increases the curves approach the $n\rightarrow\infty$ prediction $p_c = 3<1/\kappa>^{-1}$ dashed magenta line. In the insets we plot the difference $\delta p\equiv p_c-  3<1/\kappa>^{-1}$ for $n=5$ (green line) to $n=10$ (red line). The color of the inset lines changes from green to red as $n$ increases. The dots in (a) are determined from the numerical simulations discussed in Sec.~\ref{sec:Collapse}.}
\end{figure*}

In Fig.~\ref{fig:minwidth} we explore the critical pressure of rings, as found by diagonalizing Eq.~\eqref{totalmatrixene} for various periodic functions $\kappa(s)$ obtained by varying $w_{\min}$, $\alpha$, $n$ and $G(x)$. 
In Fig.~\ref{fig:minwidth}(a) we use a sinusoidal defect profile (as shown in the first line of Table~\ref{tab:table1}). The rings displayed as insets depict the $n=3$ case. We vary $w_{\min}$ (and through it the amplitude of the modulation) and calculate the critical pressure at which the ring becomes unstable. In our units, $w_{\min}=1$ corresponds to a uniform ring, with $p_c=3$, in all cases. The case $n=1$ to $n=5$ are shown. In Fig.~\ref{fig:minwidth}(b) we use a gaussian type defect profile, with $m=2$ and $\alpha = \pi/2$. In Fig.~\ref{fig:minwidth}(c) we again use a gaussian type profile, but with a much narrower defect width $\alpha = \pi/8$. Finally in (d) we plot a gaussian like function with a high enough $m$ value ($m=12$)  so that the modulation is almost step like, and $\alpha = \pi/2$.

A striking observation, as seen in Fig.~\ref{fig:minwidth} (and in all other cases examined, not shown here), is that the $n=4$ ring is particularly susceptible to collapse. As we discuss in Section~\ref{sec:Collapse}, the buckling of the ring is dominated by the $\dth_2$ mode. The four equispaced defects couple strongly with this ellipsoidal mode of buckling, and the orientation of the elliptical deformation is such that the weakened points of the ring coincide with the four areas of the ellipse where the curvature is maximally different than the preferred spontaneous curvature of the ring.

We can quantify this argument within perturbation theory.
For the constant bending modulus, unperturbed system the matrix $\hat{E}$ is diagonal,  $E_{mq}=\delta_{mp}\epsilon_m$. Since $\epsilon_m=\epsilon_{-m}$, all
 the eigenstates are doubly degenerate. We can use the two eigenstates to construct two linearly independent states, one purely real and one purely imaginary, that satisfy $\dth_m^*=\dth_{-m}$.
Because of this degeneracy, we cannot straightforwardly use perturbation theory to predict the behavior at small $\delta\kappa(s)$. Instead, as the unperturbed state we will consider the diagonal matrix plus the matrix elements at -2,2 and 2,-2 (the numbering of the matrix elements follows that of the $\dth_m$ modes, as in Eq.~\eqref{bendmels}). That matrix can be straightforwardly diagonalized, and the two lowest eigenvalues now are $\lambda_1 = 4(\kappa_0-p/3-|\ka_4|)$ and $\lambda_2 = 4(\kappa_0-p/3+|\ka_4|)$. Note that a nonzero $\ka_4$ will cause $\lambda_1$ to go negative before $\lambda_2$, and will lower the critical buckling pressure. Upon assuming that all other $\ka_m$, $m\ne \pm 4$, are small, we use second order perturbation theory in the remaining off-diagonal matrix elements to obtain:
\begin{eqnarray}\label{eigenapprox}
\lambda_1 &=& 4(\kappa_0-p/3-|\ka_4|)+ \nonumber\\
&{}&4\sum_{k>2}  \frac{k^2\;|\ka_{k-2}-\ka_{k+2}e^{-i\phi}|^2}{4(\kappa_0-\frac{p}{3}-|\ka_4|)-k^2(\kappa_0-\frac{p}{k^2-1})}+O(\ka_k^4)\nonumber\\
&&
\end{eqnarray}
where $e^{-i\phi}=\ka_{-4}/|\ka_4|$.

Eq.~\eqref{eigenapprox} cannot be solved analytically, but we can iteratively approximate:
\begin{equation}\label{pcapprox}
p_c \simeq  3(\kappa_0-|\ka_4|) - 3\sum_{k>2}  \frac{ |\ka_{k-2}-\ka_{k+2}e^{-i\phi}|^2 }{ \kappa_0 - 3\frac{\kappa_0-|\ka_4|} {k^2-1} }.
\end{equation}
This approximation is only accurate for $n=4$ rings, since for n=4, the dominant $\kappa(s)$ Fourier coefficient is $\ka_4$ and the above diagonalization captures the behavior of the system even for relatively large modulation amplitudes. We plot Eq.~\eqref{pcapprox} in Fig.~\ref{fig:minwidth} (dashed cyan line), and observe excellent agreement with the numerical results for $w_{\min}$ close to $1$. 

The $p_c(w_{\min})$ function can be calculated analytically in the $n\rightarrow \infty$  limit of very rapid periodic variation in $\kappa(s)$ if one considers the nonlinear differential equations for a general $\kappa(s)$ as derived by a functional minimization that leads to a generalization of Eq.~\eqref{small_nonunieq} (see App.~\ref{appeqderiv}), namely
\begin{eqnarray}\label{bucklingequationkappa}
\frac{\dd^2 [\ka(s)\;(c-1))]}{\dd s^2}&+&c\int^s \dd\xi \left[\ka(s)'\; c (c-1) + \ka(s)\; c'c\right]  \nonumber\\
&&\;\;\;\;\;\;\;\;\;\;\;\;\;\;\;\;\;\;\;\;\;- p -bc = 0.
\end{eqnarray}
Upon expanding Eq.~\eqref{bucklingequationkappa} to linear order in $\Delta c=c-1$ we have a generalization of Eq.~\eqref{small_nonunieq_linear}, namely,
\begin{equation}\label{bucklingequationlinearkappa}
\frac{\dd^2 \left[\ka(s) \Delta c\right]}{\dd s^2}+\left(1-\frac{b}{\ka(s)} \right)\ka(s) \Delta c = (b+p).
\end{equation}
The integration constant $b$ is now set equal to $-p$, so that the constant term  that  appears in Eq.~\eqref{bucklingequationlinearkappa} is zero.
In the limit that $1/\ka(s)$ varies much faster than $\ka(s)\Delta c$, i.e. when $n\gg 1$ or when the ring buckling is dominated by the $\dth_2$ mode, we can replace $1/\kappa$ in the parenthesis by its average:
\begin{equation}\label{averagekappa}
 \langle \frac{1}{\kappa} \rangle \equiv \frac{n}{2\pi} \int\limits_{-\pi/n}^{\pi/n}\frac{1}{\kappa} \dd s. 
\end{equation}

\begin{figure}
\includegraphics[scale=0.80]{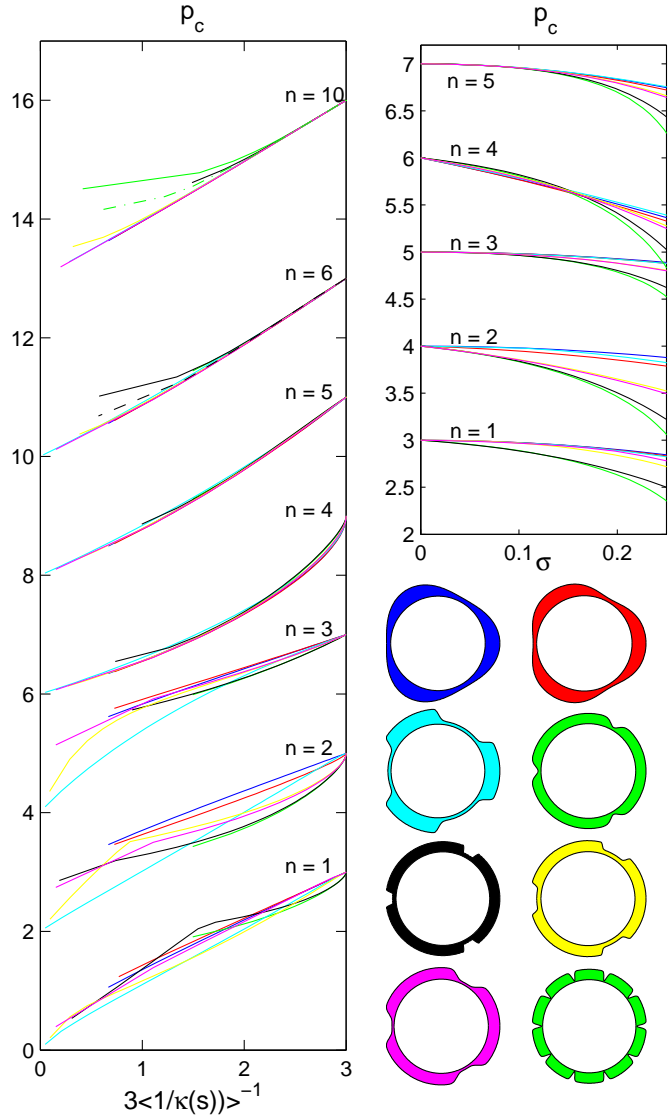}
\caption{\label{fig:oneoverkappa}  Left: $p_c$ plotted versus $3\la1/\ka \ra^{-1}$ for different ring profiles and different $n$. In our units, $\la 1/\ka (s)\ra = 1$ represents a uniform ring. The  discretization matrix cutoff is $M=300$. The $n=3$ bending modulus profile of the ring is presented in the bottom right of the graph, the colors of each ring corresponding to the colors of each plot on the left. For $n\ge 4$, we observe that all profiles collapse on the same $n$-dependent curve. The dashed curves corresponding to $n=6$ and $n=10$ are derived with a matrix cutoff $M=500$.  For better visualization each curve is displaced by $\Delta p = 2$ ($\Delta p = 3$ for $n=10$) from the curve below. Right: Plot of $p_c$ versus $\sigma$ for $n=1-5$. (Color online.)}
\end{figure}

The dashed magenta line in Fig.~\ref{fig:oneoverkappa}(a)-(d) is the $p_c(n\rightarrow\infty) = 3\langle1/\kappa\rangle^{-1}$.  Indeed, we do observe that for increasing $n$ the curves approach the limiting value  $p_c(n\rightarrow\infty) $. In the inset we plot the difference $\delta p
\equiv p_c(n\rightarrow\infty)-p_c(n) $ for $n=5-10$. The green curve denotes the $n=5$ plot and the red the $n=10$ case, and the color progresses from green to red as $n$ increases. Note that the difference for each value of $w_{min}$ monotonically approaches 0 for large $n$, as expected. 

When the functional form of $G(x)$ allows it, $3\la 1/\kappa \ra^{-1}$ can be computed analytically. For example, for a sinusoidally modulating $G(x)$ we have: 
\begin{equation}\label{oneoverkappasin}
3\la 1/\kappa\ra^{-1} = 3 \sqrt{w_{\min}(2\kappa_0-w_{\min})}.
\end{equation}

As evident from Fig.~\ref{fig:oneoverkappa}, for high $n$ the critical pressure only depends on the mean of the inverse bending modulus. Remarkably, plotting $p_c$  versus $3\la 1/\kappa\ra^{-1} $ for different $n$ in Fig.~\ref{fig:oneoverkappa}, reveals that the lines for all defect widths $\alpha$ and shapes $G(x)$ to a very good approximation collapse on the same $n$-dependent curve for $n\ge 4$. This means that the critical pressure is only a function of $\la 1/\kappa\ra$ and $n$, regardless of the detailed shape of the modulation of $\ka(s)$. As $n$ increases, $p_c(n\rightarrow\infty)=3\la 1/\kappa\ra^{-1} $ as discussed earlier. The plots for narrow defects, large $n$ and small $3\la 1/\kappa\ra^{-1} $ (corresponding to deep notches in the ring) deviate from the expected curve. This behavior is due to the finite cutoff $M$ that we have chosen to numerically compute the eigenvalues of $\hat{B}=\hat{K}\hat{E}_{\mrben}\hat{K}$. This cutoff limits the accuracy of this method for abruptly changing $\ka (s)$. It implies a discretization of the arc-length, and when $M$ is increased, the discretized system describes better the continuum case. For narrow and deep defects a finer discretization is required to represent the behavior of a continuous ring. Indeed, the plotted dashed lines correspond to $M=500$, versus $M=300$ for the rest of the plots in Fig.~ \ref{fig:oneoverkappa}, and we observe that the agreement with the master $3\la 1/\kappa\ra^{-1}$ dependent curve becomes better.

Because $\ka(s)$ varies too abruptly, the truncated matrix diagonalization approach does not allow us to examine hinged rings, like the ones of Refs.~\cite{Wang1985, Wang1993}. However, despite the different regime, the qualitative result that the $n=3$ periodic ring has the highest stability still holds for hinged rings.

Since $\kappa(s) = \kappa_0 + \dk(s)$, for $\dk(s)\ll \kappa_0$, we have 
\begin{equation}\label{oneoverkappaapproxstd} 
\la\frac{1}{\kappa}\ra^{-1}  \simeq \frac{1}{\kappa_0} \left( 1+\frac{\sigma^2}{\kappa_0^2}\right), 
\end{equation}
where $\sigma^2 = \la \dk^2(s)\ra$ is the standard derivation of bending rigidity variation, and $\kappa_0$ is defined such that $\la \ka (s)\ra = 0$. On the right of Fig.~\ref{fig:oneoverkappa} we plot $p_c$ versus $\sigma$, for small $\sigma$, for $n=1-5$. The curves for $n=1$ and $n=2$ do not agree even for $\sigma\ll 1$. However, we note that most of the discrepancy is explained by the difference in the defect width $\alpha$. For example, the black and green curves ($\alpha = \pi/8$) collapse for small $\sigma$. For small defect number $n$, when the actual defect size $\alpha/n$ is a considerable percentage of the ring circumference, and the averaged quantity $\la 1/\ka\ra$ is not sufficient for predicting the ring stability by itself. The functional form $G(x)$ of the defect has little effect on the critical pressure, at least for small variances $\sigma$.

\subsection{Rings with a randomly modulated bending rigidity}

\begin{figure}
\includegraphics[scale=0.55]{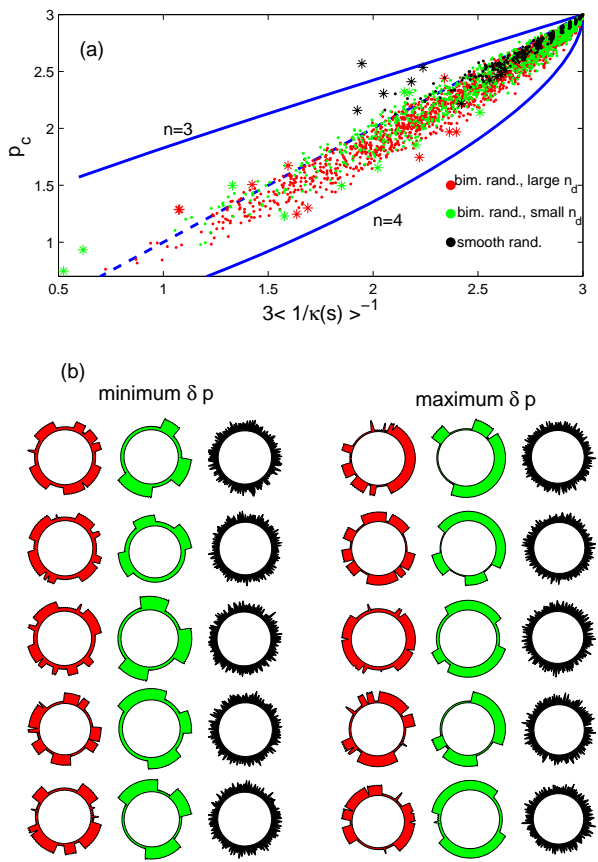}
\caption{\label{fig:randomrigsA} Critical pressure of rings with quenched randomness in the bending rigidity $\kappa(s)$. (a) Critical pressure plotted versus $p_c(n\rightarrow \infty) = 3\langle 1/\kappa \rangle^{-1}$ for a large variety of random rings . Red: bimodal randomness, large number of weakened regions. Green: bimodal randomness , small number of weakened regions. Black: Smooth randomness. The five asterisks indicate the points which deviate the most from the prediction $p_c=3\langle 1/\kappa \rangle^{-1}$, denoted with the dashed blue line. The higher blue line is the $n=3$ sinusoidal modulating $\kappa(s)$ curve and the lower blue line is the corresponding $n=4$ case. The left three columns of the plotted rings are the five realizations with the minimum  $\delta p = p_c-3\langle 1/\kappa \rangle^{-1}$, plotted in ascending order, and the right three columns the five realizations with the maximum  $\delta p$, plotted in descending order. (Color online)  }
\end{figure}

\begin{figure}
\includegraphics[scale=0.8]{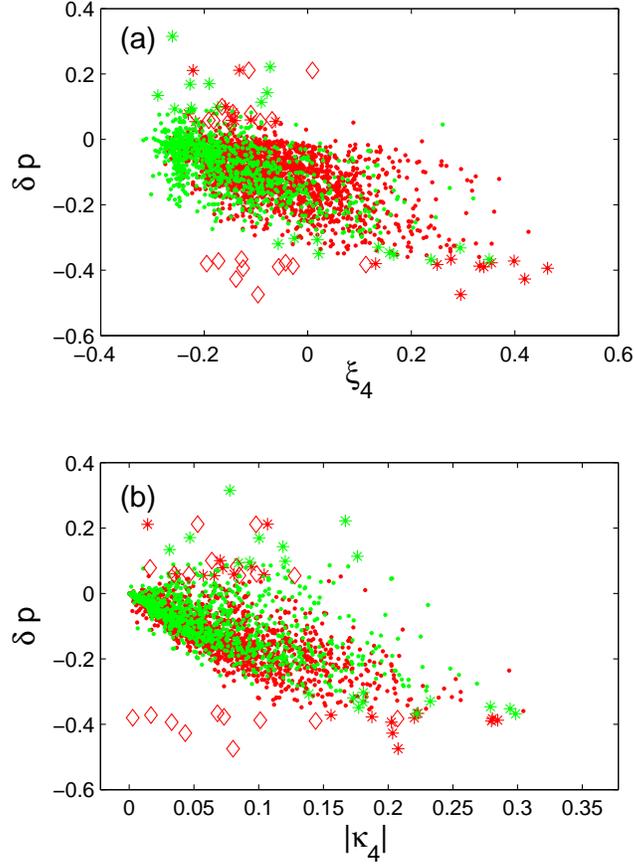}
\caption{\label{fig:randomrigsB} Critical pressure of rings with quenched randomness in the bending rigidity $\kappa(s)$. red: bimodal randomness, large number of weakened regions ($n_d\simeq 7-10$). green: bimodal randomness , small number of weakened regions ($n_d\simeq 2-3$). The asterisks indicate the points which deviate the most from the prediction $p_c=3\langle 1/\kappa \rangle^{-1}$. (a) Scatter plot of $\delta p$ versus $\xi_4$ for rings with step randomness.  The stars indicate the minima and maxima for the two cases. For comparison, the diamonds indicate the ten minima and maxima for $\xi_5$. (b) Scatter plot of $\delta p$ versus $|\ka_4|$ for rings with bimodal randomness.  The stars indicate the minima and maxima for the two cases. For comparison, the diamonds indicate the ten minima and maxima for $|\ka_5|$. (Color online)}
\end{figure}

Finally, we discuss rings with quenched randomness in $\kappa(s)$. For definiteness, we consider first rings with bimodal stepwise randomness in the bending modulus. Thus, $\kappa(s)$ can take only two values, $w_{\min}$ and $w_{\max}$, with transitions at a number $n_d$ of arbitrarily chosen points at positions $(s_1, s_2, s_3,\dots, s_{n_d})$. We then pass $\kappa(s)$ through a low pass filter to ensure that $\ka_m=0$ for $m\ge M$ and our continuous elasticity formalism is valid. However, our filter allows for cases when $\ka(s)$ is arbitrarily small at isolated regions. In this case, the value of $3\langle 1/\kappa \rangle^{-1}$ becomes inaccurate, skewed toward smaller values. The outlying positions of the five black rings with maximum $\delta p$ in Fig.~\ref{fig:randomrigsA}(a) is due to such effects.

  We also generate rings with more continuous quenched randomness by discretizing the arc-length to $N\simeq 600$ points, and assigning a random $\kappa(s_n)$ (chosen from a box distribution) to each of them (we control the average spread of $\dk(s)$ in the box distribution so that we can vary $\sigma$). We again pass $\kappa(s)$ through a low pass filter to generate smoother versions of these more continuously random rings. We then determine the critical pressure for particular realizations by matrix diagonalization, as we did for periodic rings. In Fig.~\ref{fig:randomrigsA}(a), the red data points correspond to the bimodal randomness with a large number of weakened regions ($n_d\simeq 7-10$), the green data to step randomness with small $n_d\simeq 2-3$, and the black to local randomness. The critical pressure for various realizations and values of $3\langle 1/\kappa \rangle^{-1}$ is plotted in Fig.~\ref{fig:randomrigsA}(a). Note that the majority of the results lie below the $p_c(n\rightarrow\infty)$ curve, denoted with the dashed blue line. For fixed average inverse bending modulus $\langle 1/\kappa\rangle$, periodicity and large $n$ in the bending modulus increases the ring stability. The lower blue curve is the $n=4$ periodic case, the upper is the $n=3$ case. We observe that the $n=4$ periodic case provides a lower limit for the critical pressure of rings with quenched randomness. To visualize the $\kappa(s)$ profile of rings with quenched randomness in  Fig.~\ref{fig:randomrigsA}(b), we plot the five cases with the maximum  $\delta p = p_c-3\langle 1/\kappa \rangle^{-1}$ in descending order (right three columns), and the five cases with the minimum  $\delta p $ in ascending order (three left columns). The rings with continuous randomness with the maximum $\delta p$ (red color, right column) include regions where the bending modulus nearly vanishes and cannot adequately represent the continuous behavior.

At first inspection, the ring bending modulus profile does not seem to explain the low stability of the left column for bimodal randomness (red and green columns). Since the limiting stability values are determined by the periodic cases, we need to describe how close is a ring to an $n$-fold periodic profile. For that purpose, we define the periodicity self correlation $\xi_n$:
 \begin{equation}\label{periodicityselfcorrel} 
\xi_n\equiv \frac{1}{(n-1)\sigma^2}\sum_{m=1}^{n-1}\langle \dk (s) \dk \left(s+\frac{2\pi m}{n} \right) \rangle
\end{equation}
This quantity is a measure of how close is the shape of a ring to being $n$-periodic. For rings with $\dk(s)=\dk(s+2\pi/n)$, it reaches its maximum value $\xi_n=1$.  
As we can see in the scatter plot Fig.~\ref{fig:randomrigsB}(a), most of the minimum $\delta p$ values lie on the bottom right part of the plot. Scatter plots for any other $n$ do not exhibit such a trend. For comparison, we plot the minimum and maximum $\delta p$ values plotted against $\xi_5$ as diamonds. We should note here, however, that the measure $\xi_4$ will be very high for an $n=8$ periodic-like ring. To distinguish between the two cases we can look at the $\ka_4$ Fourier mode as a measure of 4-periodicity. In Fig.~\ref{fig:randomrigsB}(b) we present the scatter plot of $\delta_p$ versus $|\ka_4|$ and we observe a similar pattern as in Fig.~\ref{fig:randomrigsB}(a).

\section{\label{sec:Collapse}Ring deformations beyond the buckling transition}

The quadratic approximation used in Sec.~\ref{sec:Criticalpress} is sufficient for determining the stability of the rings, but cannot provide information about the post-buckling behavior of the rings, as the higher order $\dth$ modes are necessary to describe the behavior of the system.
In this section we briefly discuss the initial stages of the post-buckling behavior of rings with a periodically modulating bending modulus, studied using computer simulations.

For the simulations we used a tethered chain model very similar to the LSF model of Ref.\cite{Leibler87}. In our case, we considered rings with spontaneous curvature $c_0=1/r_c$ so that the $p=0$ shape is stress free and circular. The discretized bending energy we simulate reads:
\begin{equation}\label{bendingenergy1discretized}
E_{\mrben}  = \frac{1 }{\Delta s} \sum\limits_{j=1}^N \kappa_j {(1 - \cos(\Delta\theta_j - \Delta\theta_0) )} 
\end{equation}
where $\Delta\theta_j$ is the angle between the links $j$ and $j-1$. $\kappa_j$ is the arc-length dependent bending rigidity characterizing the region between $s_j$ and $s_{j-1}$. The spontaneous curvature enters Eq.~\eqref{bendingenergy1discretized} through $\Delta\theta_0$ which is the zero bending energy angle between two successive links. For $c_0=1/r_c$, $\Delta\theta_0= 2\pi /N$. Here, $\Delta s$ is the tether length, and when $N$ is large,  $\Delta s\simeq 2\pi r_c/N$. The pressure term is as in Eq.~\eqref{volumeenergy1}, only now the area is calculated as the area of an $N$-sided polygon. For computational convenience, we did not strictly impose inextensibility to the tethers, but instead allowed them to stretch while imposing a very high energetic cost to deviations of the link length from the desired length $\Delta s$:
\begin{equation}\label{stretchingenergy1discretized}
E_{\mathrm{str}}= Y_{\mathrm{str}} \sum\limits_{j=1}^N (| \mathbf{r}_{j+1}- \mathbf{r}_j | - \Delta s )^2 
\end{equation}
with the stretching modulus chosen such that $Y_{\mathrm{str}}\gg  \frac{\min{\kappa} }{ \Delta s} $.

\begin{figure}
\includegraphics[scale=0.75]{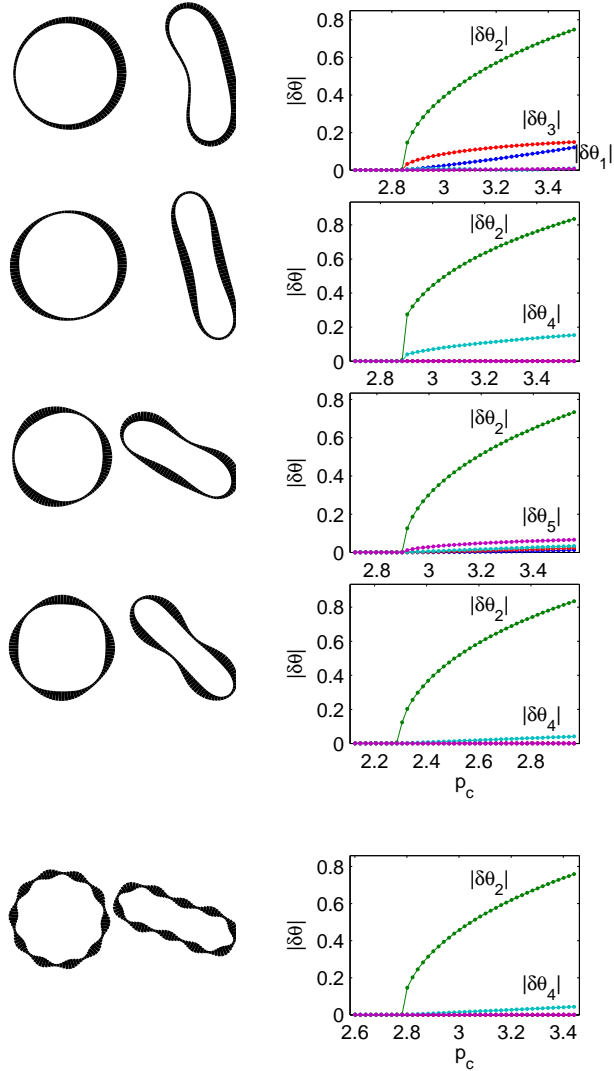}
\caption{\label{fig:postbuckling} Left: Rings with periodic sinusoidal bending rigidity profiles above and below the critical buckling pressure. On the right of each ring pair, we plot the lowest nonzero $\dth_m$ amplitude versus the applied hydrostatic pressure $p$. Blue line:  $|\dth_1|$, green line:  $|\dth_2|$, red line:  $|\dth_3|$, cyan line:  $|\dth_4|$, magenta line:  $|\dth_{5}|$. Note the lower buckling threshold for $m=4$. (Color online)  }
\end{figure}

Using a simple adaptive step gradient descent method we determined the minimum energy ring configuration for different sinusoidally modulated bending moduli choices $\{\kappa_j\}$ and pressures $p$. At each pressure $p$, we decomposed the minimum energy ring profile into $\dth$ modes. We plot some representative cases in Fig.~\ref{fig:postbuckling}. At the left column, we plot the pre and post buckling shape for $n=1-4$ and $n=10$ rings, accompanied by the amplitude of the relevant $\dth_m$ modes as pressure increases, on the right. The post buckling shape is shown for the maximum pressure displayed on the right. In all cases plotted $\Delta w = 0.8$.
The $\dth_m$ modes are zero in each case below a certain critical pressure, and acquire a non zero value above that. Due to finite size effects, the second order phase transition is rounded out slightly, and determination of the critical pressure from the graphs becomes harder. Critical slowing down makes the simulations converge slowly near the critical pressure, thus making the data in this region less reliable. Hence, to estimate the value of $p_c$ from the data, 
we fitted a $2$nd order polynomial to the post-buckling data $|\dth_2|$and extrapolated the fit to $|\dth_2|=0$. The critical values extracted this way are shown, for several $\Delta w$, in Fig.~\ref{fig:minwidth}(a), and were discussed in Section~\ref{stabilityofthering}. We observe excellent agreement between the numerical data and the eigenvalue method.

\begin{figure}
\includegraphics[scale=0.33]{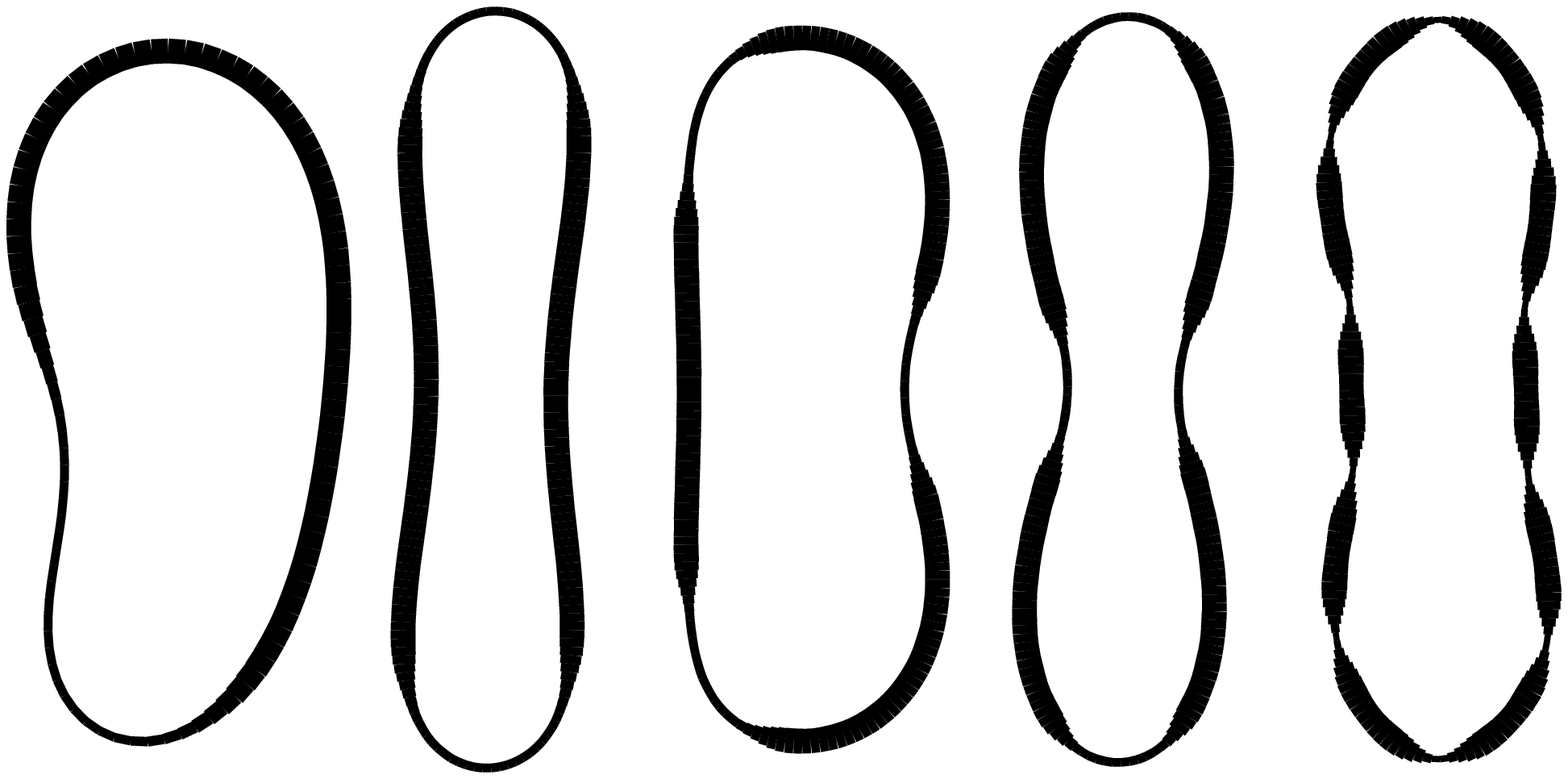}
\caption{\label{fig:postbuckling2} Post buckling states of rings with abrupt step-like modulation and $\alpha = \pi/3$. The ellipsoidal mode of deformation dominates.   }
\end{figure}

The approximation that the $\dth_1$ ring amplitude can be neglected is still valid after the buckling transition. $\dth_1$ is always much smaller than the other modes that dominate the buckling, especially close to the critical point. In particular, regardless of the bending modulus profile, the $\dth_2$ mode always dominates, driving the transition. The rings we have studied always buckle into an ellipsoidal shape. As an example of non-sinusoidally modulating ring, in Fig.~\ref{fig:postbuckling2} we plot the initial post-buckling shapes of rings with a gaussian type, ($m=8$, $\alpha=\pi/3$) bending modulus, and observe that again the $\dth_2$ mode dominates.

\begin{acknowledgments}
This work was supported by the National Science Foundation, through the Harvard Material Research Science and Engineering Center, through Grant DMR-0213805 and through Grant DMR-0654191. EK would like to thank Jacques Dumais and Jose Bico for useful discussions and suggestions.
\end{acknowledgments}

\appendix

\section{Derivation of the curvature equation for pressurized rings}\label{appeqderiv}

For completeness, in this Appendix we generalize the derivation found in Ref.~\cite{Tadjbakhsh67} for $\ka(s)\ne$const and derive the non-linear differential equation for the curvature of the hydrostatically pressurized ring with an arc-length dependent bending modulus. 

The tangent angle $\theta(s)$ of the ring that minimizes the sum of the bending energy and the pressure term subject to the ring closure constraint is found by functional minimization of:
\begin{eqnarray}\label{appEtheta}
E_{t}[\theta(s)]&\equiv&\int\limits_{ - \pi}^{\pi}  {\dd s\; \kappa(s) \left[\frac{d\theta(s)}{\dd s} - c_0\right]}^2\nonumber\\
&+&\frac{p}{2}\int\limits_{-\pi}^{\pi} {\dd s} \int\limits_{-\pi}^s {\dd\xi \sin \left[ {\theta (s) - \theta (\xi )} \right]} \nonumber\\
&+&\lambda_1\int\limits_{-\pi}^{\pi}\dd s\cos\theta(s)+\lambda_2\int\limits_{-\pi}^{\pi}\dd s\sin\theta(s).
\end{eqnarray}
We have set the radius of the ring $r_c=1$.
The first term is the bending energy $E_{\mrben}$, the second term is the work done by the pressure $E_{\mrpr}$ and the last two terms are the Lagrange multiplier constraints that come from the real and imaginary parts of Eq.~\eqref{constraint1}. The functional derivative $\frac{\delta E_t[\theta(s)]}{\delta[\theta(s)]}$ of Eq.~\eqref{appEtheta} if zero for the energy minimizing $\theta(s)$:
\begin{widetext}
\begin{eqnarray}\label{appfunderEtheta}
\frac{\delta E_t[\theta(s)]}{\delta[\theta(s)]}=& - &\left[\ka(s)\left( \theta'(s) -c_0   \right)   \right]' 
+ p \int_{-\pi}^s d\xi \cos(\theta(s)-\theta(\xi))\nonumber\\
&-& \frac{p}{2} \int_{-\pi}^{\pi} d\xi \cos(\theta(s)-\theta(\xi))-\lambda_1\sin(\theta(s))+\lambda_2\cos(\theta(s))=0
\end{eqnarray}
\end{widetext}
where the prime stands for differentiation with respect to the arc-length.
To eliminate the constants $\lambda_1$ and $\lambda_2$ we apply the operator $\mathcal{L} = \frac{1}{\theta'(s)}\frac{d}{ds}\left[\frac{1}{\theta'(s)} \frac{d}{ds}  \right]$ to both sides of Eq.~\eqref{appfunderEtheta}. After some algebra, it is easy to see that the sum $\mathcal{L}\left[\frac{\delta E_t[\theta(s)]}{\delta[\theta(s)]}\right]+\frac{\delta E_t[\theta(s)]}{\delta[\theta(s)]}$ is independent of $\lambda_1$ and $\lambda_2$ and equal to:
\begin{eqnarray}\label{appcurv}
&&\mathcal{L}\left[\frac{\delta E_t[\theta(s)]}{\delta[\theta(s)]}\right]+\frac{\delta E_t[\theta(s)]}{\delta[\theta(s)]}=\nonumber\\
&&\;\;\;\;\;\;\;\;-\frac{1}{c(s)}\left[  \frac{ \left[\kappa(s) (c(s)-c_0)  \right]'' }{c(s)}  \right]'   -  \left[\kappa(s) (c(s)-c_0)  \right]' \nonumber\\
&&\;\;\;\;\;\;\;\; \;\;\;\;\;\;\;\;  \;\;\;\;\;\;\;\; \;\;\;\;\;\;\;\; \;\;\;\;\;\;\;\;  +   \frac{p}{c(s)} \left[ \frac{1}{c(s)}    \right]'= 0
\end{eqnarray}
where $c(s)\equiv\frac{d\theta}{ds}$.

Eq.~\eqref{bucklingequation} can be easily derived from the general Eq.~\eqref{appcurv}, by setting $\ka(s)=\ka=$const, multiplying the equation by $c(s)$ and integrating once with respect to $s$. The constant $b$ in Eq.~\eqref{bucklingequation} comes from the integration. Note that since in Eq.~\eqref{appcurv} the spontaneous curvature $c_0$ appears only coupled to derivatives of $\kappa(s)$, it drops out of the equation when the bending modulus is a constant.

Similarly, Eq.~\eqref{bucklingequationkappa} comes from Eq.~\eqref{appcurv} by setting $c_0=1$,  multiplying by $c(s)$ and integrating with respect to $s$.  Eq.~\eqref{small_nonunieq} follows immediately from Eq.~\eqref{appcurv} if we set $p=0$.

\section{Tangent angle of the equilibrium zero pressure shape}\label{appthetanop}

A particular solution of the non-homogeneous Eq.~\eqref{small_nonunieq_linear} can be easily determined by inspection to be $\Delta c_1=(c_0-1)\dk/\kappa_0  + b/\kappa_0$. Since $\Delta c_1=\frac{\dd \theta(s)}{\dd s}-1$ needs to integrate to zero to be a valid solution if $\theta(s)$ is periodic, we find that $b=0$, so $\Delta c_1(s) \propto \delta\kappa(s)$. Adding the solution to the homogeneous equation, the curvature of the ring finally reads:
\begin{equation}\label{appcurvnonuni}
c(s) = -\alpha_1\sin s + \alpha_2\cos s + \frac{c_0-1}{\kappa_0}\delta\kappa +1,
\end{equation}
where $\alpha_1$ and $\alpha_2$ are $s$-independent constants.
The tangent angle $\theta(s)$ of Eq.~\eqref{non_uni_thetas} can be easily evaluated by integrating Eq.~\eqref{appcurvnonuni} with respect to $s$.
$\theta(s)$ will not satisfy the closure constraint Eq.~\eqref{constraint1} for arbitrary constants $\alpha_1$ and $\alpha_2$. To determine these constants we expand  Eq.~\eqref{constraint1} to lowest order in $\theta(s)-(s+\theta_0)$
\begin{equation}
\int\limits_{-\pi}^{\pi}\left[\alpha_1 \cos s+\alpha_2 \sin s+ \frac{(c_0-1)}{\kappa_0} \int\limits_{-\pi}^s \dd\xi \dk(\xi)  \right]e^{is}\dd s\simeq 0,
\end{equation}
and solve the real and imaginary part of the equation for $\alpha_1$ and $\alpha_2$ respectively.

\section{Buckling of rings and second order phase transitions}\label{ap_phasetrans}

In this Appendix we discuss how the buckling of pressurized rings can be viewed in the context of the Landau phenomenological theory of phase transitions.

It is known that the buckling of a uniform, hydrostatically pressurized, inextensible ring is a subcritical pitchfork bifurcation \cite{Chaskalovic95, Atanackovic98}. The stable, circular state becomes unstable above the critical pressure $p_c$, where the ellipsoidal solution branch emerges in the bifurcation diagram.  The ellipsoidal solution continuously becomes the circular solution as $p\rightarrow p_c^+$, a behavior reminiscent of a second order phase transition, with pressure being the control parameter.

The energy of the ring, when $\ka(s) = \mathrm{const.} = \ka_0$ and $\theta (s) = s + \dth (s)$, reads
\begin{eqnarray}\label{app2phasetransenergy}
&&E[\dth(s)]= \frac{\ka_0}{2}\int\limits_{ - \pi }^{\pi}  {\dd s \left(\frac{\dd\dth(s)}{\dd s}\right)}^2\;\;\;\;\;\;\;\;\;\; \nonumber\\
&&\;\;\;\;\;+  \frac{p}{2}\int\limits_{-\pi}^{\pi} {\dd s} \int\limits_{-\pi}^s {\dd\xi \sin \left[s-\xi+ {\dth (s) - \dth (\xi )} \right]}.
\end{eqnarray}
This quantity is similar to the free energy of a one dimensional spin model $\bm{m}(s)=[\cos\dth(s), \sin\dth(s)]$, where the spins are arranged on a line with periodic boundary conditions.   
However, we should proceed with this analogy with caution. The interaction of the spins $\bm{m}(s)$ in the second term is non-local, with the spins at positions $s$ and $s+\pi$ being coupled. Another complication is that the ring closure constraint Eq.~\eqref{constraint1} should be included in the free energy Eq.~\eqref{app2phasetransenergy} with two Lagrange multipliers, one for the real and one for the imaginary part. The modified free energy is Eq.~\eqref{appEtheta} with $\dth(s)=\theta(s)-s-const$. To determine the value of the Lagrange multipliers $\lambda_1$ and $\lambda_2$ and evaluate the free energy density functional, one could solve Eq.~\eqref{appcurv} and substitute the solution into Eq.~\eqref{appfunderEtheta}, but this is not the approach we will follow here. Instead, we will use the expansion of $\dth(s)$ in Fourier modes, discussed in Sec.~\ref{sec:Criticalpress}, and implement the closure constraint through discarding the $\dth_1$ Fourier mode of the expansion.

We first discuss the important lengthscales and parameters that control the physics of the problem. In this analysis, the temperature $T$ of the system is zero. There are no thermal fluctuations or entropy terms and the configuration of the ring always is at a local energetic minimum. The thermal persistence length $\ell\sim\ka_0/k_B T$ is infinite, the definition of a correlation function in the absence of temperature becomes problematic, and there is no obvious diverging lengthscale. 
However, from the bending modulus $\ka_0$ and the pressure $p$ we can construct a ``bending'' lengthscale $\ell_b(p)\equiv(\kappa_0/p)^{1/3}$ that is of relevance to this zero temperature system. This ``bending length'', divided by the ring radius $r_c$ can be thought of as a measure of the importance of the bending term of the energy compared to the pressure term. For $p$ close to $p_c$, $\ell_b(p)\sim r_c$, but for high pressures $\ell_b(p)\ll r_c$, indicating that the ring can have equilibrium states that are sharply bent \footnote{As discussed in Refs.~\cite{Tadjbakhsh67, Carrier47}, as the pressure keeps increasing above the critical pressure, secondary local minima (of energy higher than the ellipsoidal mode global minimum) appear in the energy landscape. The ring configurations that correspond to these minima have rotational symmetry with period higher than the $n=2$ ellipsoidal deformation.}. 
Note that for \textit{high} temperatures, where $\ell(T)\ll\ell_b$, the (self-avoiding) ring will behave like a branched polymer \cite{Leibler89, Leibler87}.

\begin{figure}
\centering
\includegraphics[scale=0.8]{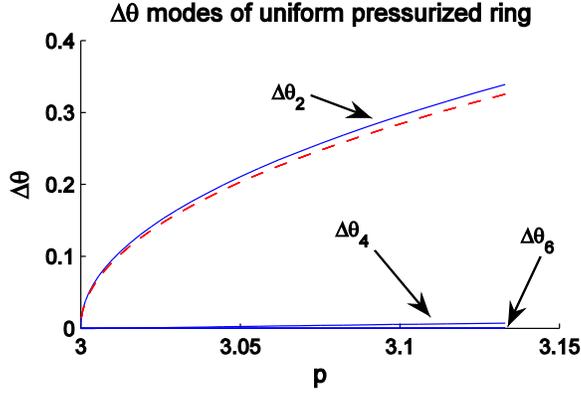}
\caption{\label{fig:ap2} Solid blue lines: Expansion of the analytical solution $\dth(s)$ of a pressurized uniform ring in $\Delta\theta_n = 2|\dth_n|$ modes. Dashed red line: Prediction of a model where only the soft mode $\dth_2$ has been taken into acount. Note the excellent agreement even far from the critical pressure $p=3$.}
\end{figure}

Only the longest wavelength (lowest bending energy) Fourier modes of $\dth(s)$ take part near the transition. This can be seen by linearizing Eq.~\eqref{bucklingequation} close to the transition point, where $\Delta c$ is small, or equivalently expanding the energy Eq.~\ref{app2phasetransenergy} to quadratic order. At $p=p_c$, as discussed in Sec.~\ref{sec:Criticalpress} the $\dth_2$ mode becomes a soft mode, and is the one driving the buckling transition. 

Having established that the soft mode $\dth_2$ is the one driving the transition, we can expand $E[\dth(s)]$ in Fourrier modes and discard all modes except $\dth_2$:
\begin{equation}\label{appexpa}
E\simeq 8\pi \left(   \left(\ka_0-\frac{p}{3}\right)  |\dth_2|^2 + \frac{4p}{15}|\dth_2|^4+\dots\right)+ p\pi
\end{equation}
where we have set $r_c=1$ for simplicity.
We see that there are no third order terms coming from the expansion of the area. This is something that we expect, since the energy should be invariant under rotations of the ring and it cannot depend on the phase of the mode $\dth_2$. Eq.~\eqref{appexpa} for $p<3\ka_0$ has a minimum at $|\dth_2|=0$, whereas for $p>p_c$ the $\dth_2=0$ solution becomes unstable. For $p-p_c\ll p_c$ the position of the minimum moves to
\begin{equation}\label{app2quad}
|\dth_2|\simeq \sqrt{ \frac{5}{8} } \left(\frac{ p-p_c}{p_c}  \right)^{1/2}.
\end{equation}
The exponent of Eq.~\eqref{app2quad} could have been easily obtained if $\dth_2$ was recognized as the order parameter of the system, and then we constructed the ``mean field'' energy Eq.~\eqref{appexpa} by expanding in $\dth_2$ and recognizing the symmetries of the system. In this sense, we can describe the buckling transition of the ring as a second order phase transition, with the pressure as the control parameter.
 In Fig.~\ref{fig:ap2} we show the expansion of the exact solution in Fourier modes (solid blue lines) and the prediction of Eq.~\eqref{appexpa} (dashed red lines). Note that the neglected $\dth_4$ and $\dth_6$ modes are very small and we observe excellent agreement  
\footnote{The energy $E[\dth(s)]$ can be analytically evaluated to infinite order in $\dth_2$ modes: $E[\dth_2] = \pi\left( 8\ka_0  |\dth_2|^2 + p \frac{\sin(4 |\dth_2|)}{4 |\dth_2|}   \right)$.}.


\end{document}